\documentclass[12pt,aps,showpacs,superscriptaddress]{revtex4}
\usepackage{epsf}
\usepackage{graphicx}
\usepackage{dcolumn}
\usepackage{amsmath}

\newcommand{\nsigma}{\mbox{\boldmath $\sigma$}}
\newcommand{\nkappa}{\mbox{\boldmath $\kappa$}}

\newcommand{\neta}{\mbox{\boldmath $\eta$}}

\newcommand{\nk}{{\bf      k}}
\newcommand{\np}{{\bf      p}}
\newcommand{\nq}{{\bf      q}}
\newcommand{\nr}{{\bf      r}}

\newcommand{\nJ}{{\bf      J}}

\begin{document}

\title{Final-state interactions and superscaling
in the semi-relativistic approach to
quasielastic electron and neutrino scattering}

\author{ J.E. Amaro}
\affiliation{Departamento de F\'{\i}sica At\'omica, Molecular y Nuclear,
          Universidad de Granada,
          18071 Granada, Spain}
\author{M.B. Barbaro}
\affiliation{
          Dipartimento di Fisica Teorica,
          Universit\`a di Torino and
          INFN, Sezione di Torino
          Via P. Giuria 1, 10125 Torino, Italy
}
\author{J. A. Caballero}
\affiliation{
Departamento de F\'{\i}sica At\'omica, Molecular y Nuclear,
          Universidad de Sevilla,
          Apdo. 1065,
          41080 Sevilla, Spain
}
\author{T.W. Donnelly}
\affiliation{
          Center for Theoretical Physics,
          Laboratory for Nuclear Science
          and Department of Physics,
          Massachusetts Institute of Technology,
          Cambridge, MA 02139, USA
}
\author{J.M. Udias}
\affiliation{
          Departamento de F\'{\i}sica At\'omica, Molecular y Nuclear,
          Universidad Complutense de Madrid,
          28040 Madrid, Spain
}

\begin{abstract}

The semi-relativistic approach to electron and neutrino quasielastic
scattering from nuclei is extended to include final-state
interactions.  Starting with the usual non-relativistic continuum
shell model, the problem is relativized by using the
semi-relativistic expansion of the current in powers of the initial
nucleon momentum and relativistic kinematics.  Two different
approaches are considered for the final-state interactions: the
Smith-Wambach 2p-2h damping model and the Dirac-equation-based
potential extracted from a relativistic mean-field plus the Darwin
factor. Using the latter the scaling properties of $(e,e')$ and
$(\nu_\mu,\mu^-)$ cross sections for intermediate momentum transfers
are investigated.

\end{abstract}


\maketitle


\section{Introduction}


In recent years, the so-called semi-relativistic (SR) approach has
been explored as a convenient and easily implementable way to
``relativize'' existing Schr\"odinger-based models of quasielastic
(QE) electron scattering from nuclei
\cite{Ama96a,Jes98,Udi99,Ama02a,Maz02,Ama03a}.  This approach
(which differs from other approaches also called
``semi-relativistic''
in the literature \cite{Bof98,Ada97}) is based on the SR expansion
of the on-shell electromagnetic current in powers of $p/m_N$ only
--- i.e., the momentum of the initial nucleon divided by the nucleon
mass --- leaving untouched the dependence on the momentum transfer
$q$ and energy transfer $\omega$.  Implementing this current
expansion and employing relativistic kinematics, the resulting SR
models can be applied at large values of the momentum transfer, of
order a GeV or more, where the traditional relativistic corrections
\cite{Fru84} based on expansions of the current in powers of $1/m_N$
are bound to fail, whatever the order of the expansion. The
existence in the literature \cite{Giu80} of expansions up to 4th
order, at least in a Foldy-Wouthuysen transformation where they are
seen to be quite complicated, raises simplicity as another advantage
of the SR approach. In fact, by carefully grouping the expansion
terms one can write them as $(q,\omega)$-dependent factors times the
traditional \cite{Sar93,Orl91} leading-order charge, convection and
spin-magnetization current operators \cite{Ama96a}. Moreover, the SR
first-order ($O(p/m_N)$) correction to the charge operator
\cite{Ama96b} is proportional to the spin-orbit charge operator
$i\vec{q}\cdot(\vec{\sigma}\times\vec{p})$, usually obtained in the
second-order $1/m_N^2$ correction in traditional non-relativistic
expansions \cite{Wal95}.

Extensive tests of the SR expansion have been performed within the
context of the Fermi gas. In particular, comparisons with the fully
relativistic Fermi gas (RFG), where the exact result is well known,
have shown the reliability of the expansion \cite{Ama96a}. Similar
SR expansions have also been performed for meson-exchange currents
\cite{Ama98a}, nucleon-$\Delta$ currents \cite{Ama99a} and more
recently for the charge changing (CC) weak current driving QE
$(\nu_l,l^-)$ reactions \cite{Ama05a}.

Part of the interest in reactions involving neutrinos instead of
electrons lies in their implications for ongoing and planned
neutrino oscillation experiments
\cite{Fuk98,Ahn03,Abl95,Chu97,Amb04,Dra04,Gra06,Zel06,Zel06b}. Since
these naturally involve heavy nuclei as targets, reliable nuclear
models of the reaction A$(\nu_l,l^-)$ play an essential role in the
interpretation of the data.  Because of the close relationships that
exist amongst all semi-leptonic electroweak processes, accurate
descriptions of A$(e,e')$ data appear as a requirement that sets
strong restrictions for nuclear modeling of neutrino reactions.
Based on what is known from previous approaches to this problem, and
on the results presented in this paper, two ingredients arise as
essential if the nuclear reaction modeling is to be successful in
describing the electroweak cross sections in the kinematical regime
of interest, namely, relativity and final-state interactions (FSI).


First is relativity, since one is dealing with momentum transfers
in
 the intermediate-to-high energy regime, typically of the order of 1 GeV or
 higher, for which
 traditional non-relativistic expansions in powers of $1/m_N$ are not
 applicable. Different fully relativistic models (based on Dirac
 equations and/or relativistic many-body theories) have been developed
 in recent years aiming to describe electron and neutrino scattering
 \cite{Pick87,Pick89,Jin92,Udi93,Udi95,Kim95,Kim95b,albe1,albe2,Chi89,Meu03,Meu04,Cris06,Mai03,Cab06}.
 While relativistic approaches based on the Dirac equation appear as
 the most direct way to deal with the problem, the importance of the
 different relativistic ingredients is more easily explored by
 detailed comparison with the extensively employed non-relativistic
 approaches, and in particular to the SR one that is the focus of this
 work.  This comparison will also help in identifying which relativistic
 ingredients (current operators, initial- or final-state wave
 functions) are the main ones responsible for the difference between
 relativistic and non-relativistic results. Upon incorporating these
 ingredients in SR approaches, one can extend their applicability
 to regions well into the relativistic domain, and thereby hope to produce reliable
 results at high energies.

Secondly, the model of FSI must account for many-body effects that
are known to be essential in describing reasonably well $(e,e')$
cross sections within the context of non-relativistic approaches
\cite{Fab89,Co88,Ama92,Pac93,Ama94,Jes94,Gil97}. For instance,
medium modifications of the one-particle one-hole (1p-1h)
final-state self-energy via two-particle two-hole (2p-2h)
intermediate states in the continuum appear to be important in
non-relativistic many-body treatments of a variety of processes
involving low-to-intermediate momentum transfers
\cite{Smi88,Bot05,Ama97}. The main effect is a significant shift
of strength to high energies in the QE cross section. On the other
hand, relativistic impulse approximation modeling of inclusive
reactions \cite{Cab05,Cab06}  shows an important effect of the FSI
where strength is shifted to the high momentum region.

The importance of FSI for intermediate-to-high momentum transfers
has also been investigated in connection with the scaling
properties of QE cross section.  The recent analysis of $(e,e')$
world data in the QE region \cite{Don99a,Don99b,Mai02} has
permitted the extraction of a universal scaling function
$f_L(\psi')$ from the longitudinal response function data (here
$\psi'$ is the dimensionless scaling variable defined in
eq.~(\ref{psi}) below). The experimental scaling function presents
an asymmetric shape, in contrast to the symmetric behavior
predicted by most independent-particle models, which essentially
give results that are similar to the RFG, with the exception of
the tails observed for $\psi'<-1$ and $\psi'>1$, where the RFG
result is zero by construction. The behavior exhibited by the
experimental scaling function reflects that of the longitudinal
response function including the FSI for intermediate values of the
momentum transfer \cite{Ama94}, i.e. the single-particle strength
at the QE peak is reduced and a large tail is observed for high
values of the energy transfer $\omega$.  Note that the difference
with respect to previous approaches is that the experimental data
now correspond to relatively high values of the momentum transfer,
mainly in the range from $q=500$ MeV/c to 1 GeV/c, for which
scaling occurs, and where most of the existing non-relativistic
models of FSI are no longer applicable.

Recently \cite{Cab06,Cab05} good agreement between the calculated
scaling function and the data has been found in a fully relativistic
approach based on the relativistic mean-field (RMF) model, within
the impulse approximation (IA). The bound and outgoing nucleon
states are described by the same self-consistent Dirac-Hartree
potential.  Here the use of the same relativistic potential in the
initial and final states appears to be essential, since, as shown in
\cite{Cab05}, calculations using the real parts of complex
relativistic optical potentials do not produce the asymmetric
behavior seen in the data. Apparently the main factors responsible
for the asymmetry are not only relativistic kinematics, but also the
particular dynamics contained in the Dirac equation when strong
scalar and vector potentials are used for the continuum nucleon
states (i.e., for the FSI). In fact, within the IA the significant
shift of strength to positive $\psi'$ values only occurs under the
presence of these strong potentials \cite{Cab06}.

In this paper we investigate whether it is possible to describe the
asymmetric behavior of the QE scaling function within the
semi-relativistic approach for intermediate-to-high momentum
transfers.  As FSI are essential in order to perform meaningful
comparisons with QE scaling data, we explore various ways of
including these ingredients in the SR model at relativistic
energies. We consider two different models of FSI: the Smith-Wambach
2p-2h damping (SWD) model and the Dirac-equation-based (DEB)
potential plus the Darwin factor (DEB+D)
\cite{Udi95,jaminon,others}.

The SWD model is an extrapolation to relativistic kinematics of
the nuclear re-interaction model introduced in \cite{Smi88,Co88},
which gives one a straightforward way to incorporate the effects
of 2p-2h excitations in the nuclear response function within a
non-relativistic context.  Speculations on how to extend this
model to QE neutrino scattering for relativistic energies have
been presented in \cite{Co01,Co02}, within the context of the RFG.

The DEB+D method attempts to translate the success of the RMF
model in describing the phenomenological scaling function from  QE
$(e,e')$ data \cite{Cab05,Cab06} to the SR approach. We perform
calculations in a SR continuum shell model where the final wave
functions are obtained by solving the Schr\"odinger equation with
the DEB potential, and multiplying by the Darwin factor.  The DEB
potential is obtained by a reduction of the Dirac equation to a
Schr\"odinger equation with a local potential, which implies that
one must multiply the wave function by a non-locality Darwin
factor \cite{Udi95,Udi01,jaminon,others}. This essentially amounts
to using upper components from the solutions of the Dirac equation
in computing the final nucleon wave functions.

Both the DEB potential and Darwin factor are functions of the
local (energy independent) vector (V) and scalar (S) components of
the relativistic Hartree potential. The non-relativistic reduction
implied in the derivation of DEB potential and Darwin term
introduces a linear dependency on the energy of the particle.  The
results obtained in the SR model describing FSI in this way
compare well with those obtained using the fully relativistic RMF
model, and thus also reproduce the successful comparison with the
experimental scaling function data. This allows us to conclude
that it is the treatment of FSI in the RMF that gives rise to the
large asymmetric tail in the superscaling function.

In this paper we also present an application of the SR-DEB+D 
approach to the superscaling analysis (SuSA) of
neutrino CC QE cross section. This method has been proposed in
\cite{Ama05b} as an efficient way of predicting neutrino cross
sections from the $(e,e')$ data, by exploiting the scaling
properties of the latter. The method is based on the hypothesis that
the neutrino cross sections universally scale in the same way as do
the electron scattering cross sections for intermediate-to-high
momentum transfers, as is seen for several types of models
\cite{Cab05,Cab06,Ama05a}. 
An application of the SuSA approach to compute integrated 
neutrino cross sections has been reported recently in \cite{Ama06}. 
In this work we investigate the validity
of the superscaling approach within the SR-DEB+D model. Given the
success of our model in describing the experimental scaling data,
this check will help further to lay the foundations of the SuSA
approach introduced originally in \cite{Ama05b}.

The structure of the work is the following. In sect.~II we briefly
outline the SR model and the different treatments of the FSI.  We
present results in sect.~III and our conclusions in sect.~IV.

\section{The Semi-Relativistic (SR) model}

In this section we summarize the basic formalism for electron and
neutrino reactions within the SR approach, including the different
treatments of the FSI. We refer the reader to \cite{Ama96a} and
\cite{Ama05a} for specific details.

\subsection{Electron and neutrino cross sections}

We focus specifically on the reactions $(e,e')$ and $(\nu_l,l^-)$
induced by electrons and neutrinos, respectively, where $l^-$ is a
lepton with mass $m_l$ (typically a muon). The four-momentum of the
incident lepton is $k^{\mu}=(\epsilon,\nk)$ while
$k'{}^{\mu}=(\epsilon',\nk')$ is the four-momentum of the final
lepton. The four-momentum transfer is denoted
$Q^{\mu}=k^{\mu}-k'{}^{\mu}= (\omega,\nq)$. The results below are
referred to a coordinate system where the $z$-axis points along
$\nq$ and the $x$-axis along $\nk-(\nk\cdot\nq)\nq/q^2$.

The inclusive cross section for these reactions can be written in general as
\begin{equation}
\frac{d\sigma}{d\Omega' d\epsilon'}=\sigma_0 {\cal F}^2  ,
\end{equation}
where $\sigma_0$ is the usual Mott cross sections for $(e,e')$ reactions
\cite{Ama02a}, while it becomes
 an analogous factor in the case of $(\nu_l,l^-)$ reactions
(see \cite{Ama05a,Ama05b} for explicit expressions). The relevant
observable is the nuclear structure function ${\cal F}^2$
which can be written as
\begin{equation}
{\cal F}^2 = v_L R_L + v_T R_T
\end{equation}
for $(e,e')$ and
\begin{equation}
{\cal F}^2 = \hat{V}_{CC} R_{CC}+2\hat{V}_{CL} R_{CL}+\hat{V}_{LL} R_{LL}
+\hat{V}_{T} R_{T}+2\hat{V}_{T'} R_{T'}
\end{equation}
for $(\nu_l,l^-)$ reactions. The lepton kinematical factors $v_K$
and $\hat{V}_K$ are defined in \cite{Ama02a} and \cite{Ama05b},
respectively.  Note that, in the limit of lepton mass $m_l=0$,
one has the identities $v_L=\hat{V}_{CC}$ and $v_T=\hat{V}_T$.
Finally, the nuclear response functions are given by
\begin{eqnarray}
R_L=R_{CC} &=& W^{00}  \label{RL} \\
R_{CL}     &=& -\frac12 \left( W^{03}+W^{30} \right)  \\
R_{LL}     &=& W^{33}  \\
R_{T}      &=& W^{11}+W^{22}  \\
R_{T'}     &=& -\frac12 \left(W^{12}-W^{21} \right)  ,
\end{eqnarray}
where the inclusive hadronic tensor is
\begin{equation}
W^{\mu\nu}=\overline{\sum_{fi}}\delta(E_f-E_i-\omega)
\langle f|J^{\mu}(Q)|i\rangle^*
\langle f|J^{\nu}(Q)|i\rangle .
\end{equation}
Here $J^{\mu}(Q)$ is the nuclear current operator relevant for the
reaction, i.e., the electromagnetic current in the case of
$(e,e')$ or the CC weak current in the case of $(\nu_l,l^-)$, as
specified below. Note that although in eq.~(\ref{RL}) $R_L$ and
$R_{CC}$ are formally equal, in practice they are not the same
quantity, since different current operators and nuclear matrix
elements are involved in their definitions, $R_L$ referring to
electron scattering and $R_{CC}$ to neutrino reactions.

The current operators used in this work are SR expansions of the
fully relativistic ones in powers of $\neta= \np/m_N$, where $\np$
is the momentum of the initial (bound) nucleon, which is typically
small ($\eta < 1/4$), and therefore an expansion at most to first
order ($O(\neta)$) should be adequate for the inclusive reactions
considered here in the region of the QE peak. The expansion was
performed in \cite{Ama96a} and \cite{Ama05a} (see \cite{Ama02a}
for a review on the general SR expansion procedure). Here we give
just the final expressions used in our calculations. The CC weak
current is written in momentum space as
$J^{\mu}=J^{\mu}_V-J^{\mu}_A$, in terms of the vector and axial
current terms. For the vector current one has
\begin{eqnarray}
J_V^0 &= & \xi_0 + i \xi'_0\ (\nkappa\times\neta)\cdot\nsigma
\label{JV0}\\
\nJ_V^\perp &=& \xi_1\ \neta^\perp + i \xi'_1\ \nsigma\times\nkappa
\label{JV}
\end{eqnarray}
and for the axial current
\begin{eqnarray}
J_A^0 &=&\zeta'_0\ \nkappa\cdot\nsigma+\zeta''_0\ \neta^\perp\cdot\nsigma \\
J_A^z &=&\zeta'_3\ \nkappa\cdot\nsigma+\zeta''_3\ \neta^\perp\cdot\nsigma \\
\nJ_A^\perp &=& \zeta'_1\ \nsigma^\perp.  \label{JA}
\end{eqnarray}
In eqs.~(\ref{JV0}--\ref{JA}) we have introduced the dimensionless
momentum transfer vector $\nkappa=\nq/2m_N$, while we use the
notation $\nJ^\perp$ to denote the component of the vector $\nJ$
perpendicular to the momentum transfer $\nq$.

Finally, the nucleon form factors and relativistic correction
factors are included in the coefficients $\xi_i$, $\xi'_i$,
$\zeta'_i$ and $\zeta''_i$ (the corresponding relativistic
versions of these quantities for the electroweak neutral current
were introduced in the appendix of \cite{Ama96a}), defined by:
\begin{eqnarray} \label{xi0}
\xi_0 = \frac{\kappa}{\sqrt{\tau}}2G_E^V
&,&
\xi'_0=\frac{2G_M^V-G_E^V}{\sqrt{1+\tau}} ,
\\ \label{xi1}
\xi'_1= 2G_M^V  \frac{\sqrt{\tau}}{\kappa}
&,&
\xi_1 = 2G_E^V  \frac{\sqrt{\tau}}{\kappa} ,
\\
\zeta'_0 =
\frac{1}{\sqrt{\tau}}\frac{\lambda}{\kappa}G_A'
&,&
\zeta''_0 =
\frac{\kappa}{\sqrt{\tau}}
\left[ G_A -\frac{\lambda^2}{\kappa^2+\kappa\sqrt{\tau(\tau+1)}}G_A'\right] ,
\\
\zeta'_3 =
\frac{1}{\sqrt{\tau}}G_A'\ \ \
&,&
\zeta''_3 =
\frac{\lambda}{\sqrt{\tau}}
\left[ G_A -\frac{\kappa}{\kappa+\sqrt{\tau(\tau+1)}}G_A'\right] ,
\\
\,\,\zeta'_1 = \sqrt{1+\tau}G_A
&.&
\label{zeta1}
\end{eqnarray}
Here use has been made of the dimensionless variables
$\kappa=q/2m_N$, $\lambda=\omega/2m_N$, and
$\tau=\kappa^2-\lambda^2$. The nucleon form factors appearing in
eqs.~(\ref{xi0}--\ref{zeta1}) are the
 isovector magnetic $G_M^V=G_M^p-G_M^n$, isovector
electric $G_E^V=G_E^p-G_E^n$ and the axial form factor $G_A$.  In
$\zeta'_i$ and $\zeta''_i$ we have introduced the following
combination of axial-vector and pseudoscalar form factors
\begin{equation} \label{gap}
G_A' = G_A-\tau G_P\ ,
\end{equation}
where $G_P$ is the pseudoscalar nucleon form factor.

The expression for the SR expansion of the electromagnetic current
is very similar to that for the vector current in
eqs.~(\ref{JV0},\ref{JV}). The only difference is that the proper
proton or neutron form factors, instead of the isovector ones,
should be inserted in the corresponding $\xi_i$ and $\xi'_i$
coefficients.

\subsection{The nuclear and FSI models}

In this work we describe the nuclear structure with an uncorrelated
shell model. The many-body initial nuclear state $|i\rangle$ is a
Slater determinant. It is  constructed with a set of single-particle
solutions of a non-relativistic mean-field potential. The final
states $|f\rangle=|ph^{-1}\rangle$ are particle-hole excitations  of
the nuclear core. The holes $|h\rangle$, with non-relativistic
energies $\epsilon_h$, are solutions of the Schr\"odinger equation
with a Woods-Saxon potential. The particles, $|p\rangle$, with
asymptotic kinetic energies $\epsilon_p=\epsilon_h+\omega$, are
continuum solutions of the Schr\"odinger equation with positive
energy. In the SR approach the relativistic kinematics is
implemented by the following substitution of the eigenvalue for
positive energies:
\begin{equation}
\epsilon_p \longrightarrow \epsilon_p\left(1+\frac{\epsilon_p}{2m_N}\right).
\end{equation}
For intermediate energies $\epsilon_h$ is small compared with
$\omega$ and the above substitution can be replaced to a good
approximation by $\lambda\rightarrow \lambda(1+\lambda)$  with small
error \cite{Alb90}.

We begin at the basic starting point, denoted below as
semi-relativistic Woods-Saxon (SR-WS), 
where the continuum states are described by solutions
of the Schr\"odinger equation with the same Woods-Saxon potential as
the bound states.
Building on this, two further different approaches to FSI have
been considered.

In a first treatment, we include the FSI effects coming from 2p-2h
intermediate states by following the Smith-Wambach damping model
(SWD). Within this model the total response function
$R_K^{SWD}(q,\omega)$ is computed from the bare SR-WS one,
$R_K^{SR-WS}$, as a folding or convolution integral
\begin{equation} \label{folding}
R^{SWD}_K(q,\omega)=\int_0^\infty dE\ R_K^{SR-WS}(q,E) \left[
\rho\left (\frac{m_N}{M^*(q)}E,\omega \right) +\rho\left
(\frac{m_N}{M^*(q)}E,-\omega \right) \right] ,
\end{equation}
where $M^*(q)$ is the nucleon effective mass taken
from \cite{Pin88} and, following \cite{Smi88}, the function
$\rho(E,\omega)$ is given by
\begin{equation}
\rho(E,\omega)
= \frac{1}{\pi}
\frac{\Gamma(\omega)/2}{\left[E-\omega-\Delta(\omega)\right]^2+
\left[\Gamma(\omega)/2\right]^2} ,
\end{equation}
\begin{equation}
\Sigma(\omega)=\Delta(\omega)+i\frac{1}{2}\Gamma(\omega)
\end{equation}
being the complex self-energy of the final states.
The imaginary self-energy is computed through the following average
for the particles and holes
\begin{equation} \label{Gamma}
\Gamma(\omega) = \frac{1}{\omega} \int_0^\omega d\epsilon
\left[\gamma_p(\epsilon)+\gamma_h(\epsilon-\omega)
\right] .
\end{equation}
We use the parametrization of \cite{Mah81}
\begin{equation} \label{gamma}
\gamma_p(\epsilon)=\gamma_h(\epsilon)=\gamma(\epsilon)=
2\alpha
\left[\frac{\epsilon^2}{\epsilon^2+\epsilon_0^2}\right]
\left[\frac{\epsilon_1^2}{\epsilon^2+\epsilon_1^2}\right]
\end{equation}
with $\alpha=10.75$ MeV, $\epsilon_0=18$ MeV and $\epsilon_1=110$ MeV.

The second approach to FSI, and main focus in this work, is the
Dirac-equation-based plus Darwin term (DEB+D). The DEB potential,
$U_{DEB}$, is obtained from the Dirac equation, written as a
second-order equation for the upper component $\psi_{up}(\nr)$
\cite{jaminon, Udi95, others}.
This wave function is then written in the form
\begin{equation} \label{psiup}
\psi_{up}(\nr)=K(r,E)\phi(\nr),
\end{equation}
and the Darwin factor $K(r,E)$ is chosen in such a way that the
function $\phi(r)$ satisfies a Schr\"odinger-like equation
\begin{equation}
\left[ -\frac{1}{2m_N}\nabla^2 + U_{DEB}(r,E) \right] \phi(\nr)=
\frac{E^2-m_N^2}{2m_N}\phi(\nr).
\end{equation}
Here $E$ is the relativistic energy of the final nucleon.
Note that both the DEB potential, $U_{DEB}(r,E)$, and the Darwin
 factor, $K(r,E)$, show an explicit energy dependence coming from the
 non-relativistic reduction of the Dirac equation.  The complete
 expressions in terms of the vector and scalar parts of the
 relativistic potential are given, for instance, in \cite{Udi95,Udi01}.
In the results section we show the energy dependence of these
quantities in more detail.

\subsection{Superscaling approach}

In this paper we focus on the properties of the scaling functions
$f_L(\psi')$ and $f_T(\psi')$: these are defined as the
electromagnetic responses $R_L(q,\omega)$ and $R_T(q,\omega)$
divided by the appropriate single-nucleon functions, $G_K$
($K=L,T$), weighted by the nucleon number involved in the process
(see \cite{Ama05b,Ama05a} for details).

The scaling functions depend generally on $q$ and $\omega$ and are
different for different nuclei. Scaling of first kind is said to
hold when they are found to depend, not independently on $q$ and
$\omega$, but only on a specific function of these two, namely the
scaling variable $\psi'(q,\omega)$ defined as
\begin{equation} \label{psi}
\psi'=
\xi_F^{-1/2}
\frac{\lambda'- \tau'}{
       \left[(1+\lambda')\tau'+\kappa\sqrt{\tau'(1+\tau')}\right]^{1/2}}
       ,
\end{equation}
where $\lambda'=(\omega-E_s)/2m_N$, $\kappa=q/2m_N$,
$\tau'=\kappa^ 2-\lambda'{}^2$, and
$\xi_F=\sqrt{1+(k_F/m_N)^2}-1$. Experimentally, the Fermi momentum
$k_F$ and the energy shift $E_s$ are empirical parameters
determined through fits to QE electron scattering data. The
unshifted scaling variable $\psi$ is defined by the same formula,
but with $E_s=0$.

If $f_L$ and $f_T$ are independent of $k_F$ as well, one says that
second-kind scaling in fulfilled. When both kinds of scaling
occur, the responses are said to ``superscale''.

In the case of neutrino reactions, scaling functions can be obtained
in a similar way, dividing the weak responses by the corresponding
single-nucleon contribution, see \cite{Ama05a,Ama05b,Cab06}.  The
approach called SuSA assumes that the scaling functions entering
both $(e,e')$ and $(\nu_l,l^-)$ reactions are the same: in this case
one can reconstruct the weak cross sections from the electromagnetic
ones by multiplying the scaling function by the weak single-nucleon
contribution.

\section{Results}

In this section we present results for the nucleus $^{12}$C. The
ground state is described here in the extreme shell model, namely
the $1s_{1/2}$ and $1p_{3/2}$ shells are fully occupied for protons
and neutrons. The Woods-Saxon potential parameters used to describe
the (non-relativistic) bound energies and wave functions are given in
 \cite{Ama05a}.  As stated in the previous section, for the final
states in the SR model we use different forms of the potential
depending on the description of the FSI. In the SR-WS the same
Woods-Saxon potential is used for initial and final states. In the
SR-DEB the DEB potential and the Darwin term are used for the
final states. A general multipole expansion of the current
operators is performed to compute the nuclear response functions
and  scaling functions.  We refer the reader to
\cite{Ama96a,Ama05a} for details on this aspect of the
calculation.

We first investigate the properties of the electromagnetic
$(e,e')$ L and T scaling functions, leaving the discussion of the
neutrino CC cross section for the end of this section.  In
particular, we are interested in the study of the properties of
scaling of the first kind.  To this end we compute the function
$f_K(\psi')$ ($K=L,T$) for fixed values of the momentum transfer
$q$. One has scaling of the first kind when no dependence on $q$
is observed.

\subsection{Relativistic effects}

We start with a brief discussion of the size of the relativistic
effects embodied in the SR model.  In Fig.~1 we show the
longitudinal and transverse scaling functions of $^{12}$C computed
in the shell model for three values of the momentum transfer,
$q=0.5$, 0.7 and 1 GeV/c. We show results for three models of the
reaction, all of them including the same Woods-Saxon potential in
initial and final states. The dashed lines correspond to a
traditional non-relativistic approach using non-relativistic
currents and non-relativistic kinematics. The dotted lines have been
obtained using relativistic kinematics, but still the
non-relativistic current operators. Finally, the solid lines
correspond to the SR approach with relativistic corrections in the
current as well, i.e., to the SR-WS model.

\begin{figure}[tp]
\begin{center}
\includegraphics[scale=0.85,  bb= 50 400 540 790]{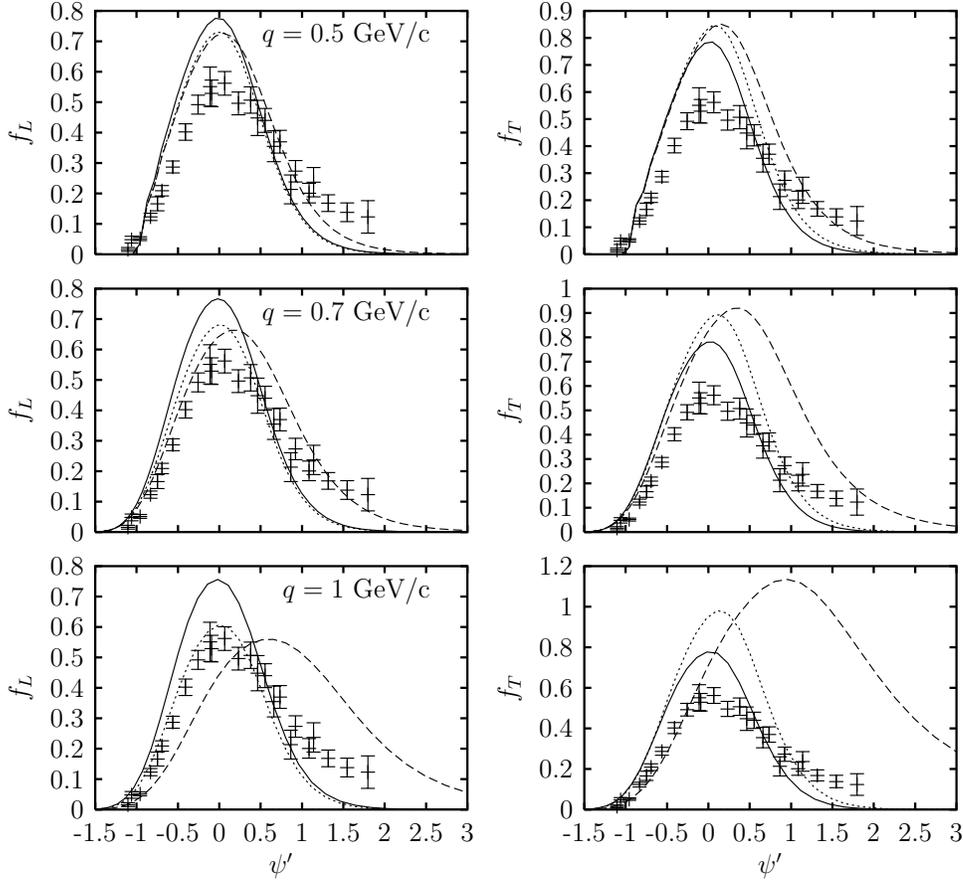}
\caption{ Scaling functions of $^{12}$C in the continuum shell model
with a Woods-Saxon potential. Solid lines: SR model. Dashed lines:
Traditional non-relativistic results. Dotted lines: Non-relativistic
current operators using relativistic kinematics. The experimental
data are taken from \cite{Mai02}. 
}
\end{center}
\end{figure}

\begin{figure}[tp]
\begin{center}
\includegraphics[scale=0.85,  bb= 50 400 540 790]{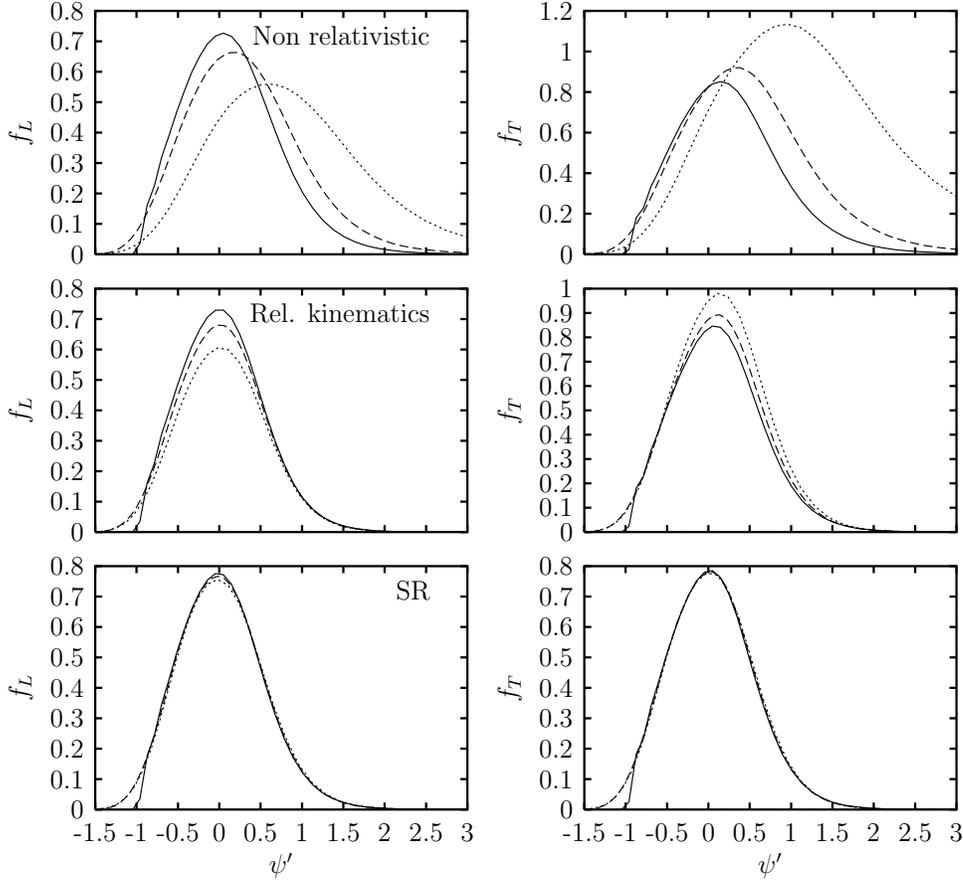}
\caption{ In each panel the $^{12}$C scaling function is displayed
for three values of $q=0.5$ (solid lines), 0.7 (dashed) and 1
GeV/c (dotted). The initial and final states are described with
the same Woods-Saxon potential. Upper panels: Traditional
non-relativistic model. Middle panels: Non-relativistic model
using relativistic kinematics. Lower panels: SR model. }
\end{center}
\end{figure}

As one can see, the width of the scaling functions is
significantly reduced in the calculations that include
relativistic kinematics, while the inclusion of relativistic
effects in the current operators leads to an increase of $f_L$ and
a reduction of $f_T$. The importance of relativistic effects
clearly increases with the momentum transfer. In the figure we
also plot the experimental data for the averaged $f_L(\psi')$,
taken from \cite{Mai02}.  It is important to point out that data
refer only to the analysis of the longitudinal scaling function,
i.e., $f_L$; however, we have chosen to present such data also in
comparison with theoretical results for $f_T$. This makes the
comparison between $f_L$ and $f_T$ more straightforward and, in
fact, it allows us to check the degree to which scaling of the
zeroth kind, i.e., $f_L=f_T$, is fulfilled for the different
models presented in this work.

Incidentally, note that the non-relativistic current with
relativistic kinematics (dotted lines) gives the best description of
experimental data for $q=1$ GeV/c, where a non-relativistic
treatment of the current operator should hardly be adequate. This
fact is a consequence of the relativistic factor
$\kappa/\sqrt{\tau}$ in the coefficient $\xi_0$ in eq.~(\ref{xi0})
which produces a reduction of $f_L$.

Although from the full set of results seen in Fig.~1 it clearly
appears that none of the three approaches provides a satisfactory
description of the experimental data, one can still study the
scaling properties of these models. One knows from previous
superscaling studies that scaling of the first kind is approximately
valid for the experimental longitudinal scaling function, $f_L$.
This places important restrictions on theoretical models, namely,
that they should be (reasonably) compatible with scaling of the
first kind. Otherwise they would be unable to describe the $(e,e')$
cross section over a wide range of momenta. This is illustrated in
Fig.~2 for the different approaches in the shell model. There we
show the same results as in Fig.~1, but presented in a different
way, namely, this time plotting together the results from each of
the models for the three values of $q$ simultaneously. From Fig.~2
one can appreciate that the SR approach is the only model that shows
scaling of the 1st kind. Therefore, we conclude that the
relativistic corrections that are consistently accounted for in the
SR, in both the current operator and the kinematics, are important
for recovering scaling of 1st kind, although the model still lacks
dynamical ingredients from relativistic FSI needed to reproduce the
shape of the experimental data, as discussed below.

\begin{figure}[tp]
\begin{center}
\includegraphics[scale=0.85,  bb= 50 400 540 790]{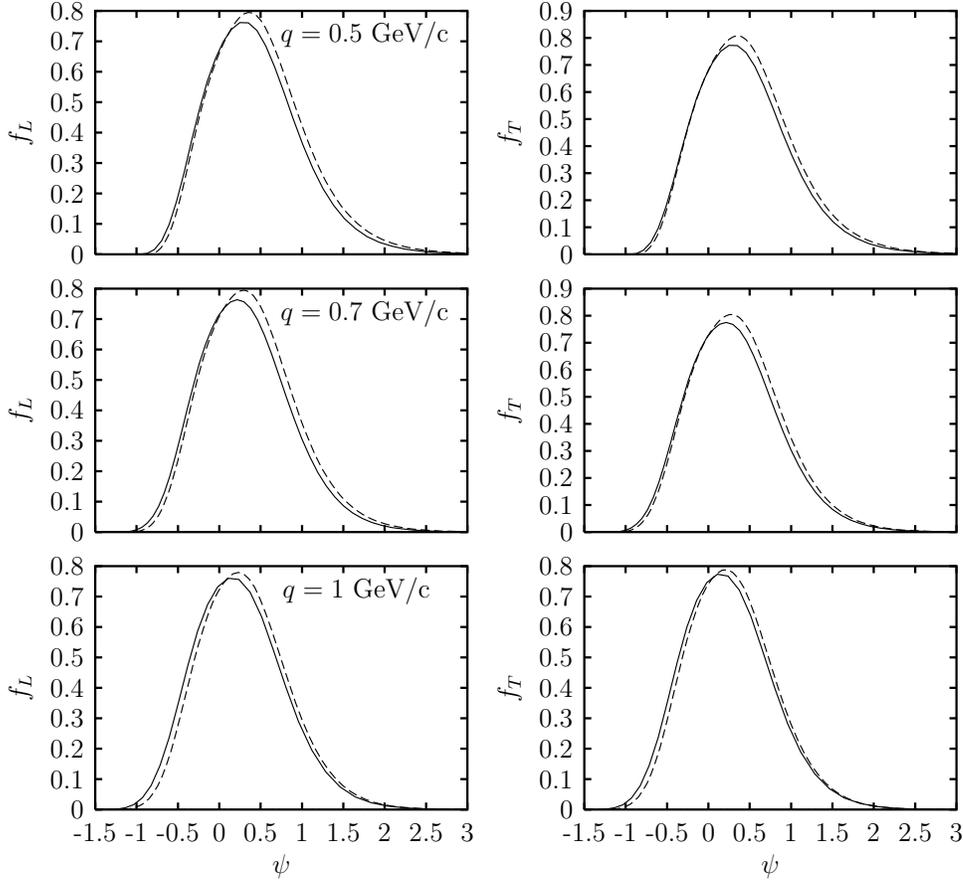}
\caption{Comparison of the semi-relativistic and fully relativistic
models in PWIA. Solid: SR. Dashed: Relativistic with the CC2 current
operator. 
}
\end{center}
\end{figure}

Before entering into a detailed discussion of FSI, in Fig.~3 we
display a test of the SR expansion, by comparing with a fully
relativistic result. Similar tests have been carried out in the
context of the relativistic Fermi gas in \cite{Ama96a,Ama05a}.
However, at the level of the SR-WS it is not possible to make
meaningful comparisons with fully relativistic results, because of
the different description of the dynamics in the final states. To
make a consistent comparison one should go a step back and consider
the plane-wave impulse approximation (PWIA), where the final
potential is set to zero. This is done in Fig.~3, where we show the
PWIA results for the L and T scaling functions in semi-relativistic
and relativistic models. From these results one can see that the
description of the initial nuclear state and of the current operator
are quite similar in both models. Note that in the relativistic
calculation the CC2 current has been employed (see also later). The
bound wave functions and energies are very close in the relativistic
and non-relativistic approaches. The small differences seen in
Fig.~3 are linked to off-shell effects in the current matrix
elements, and are of the same order as can be found between
relativistic results using different prescriptions for the current
operator \cite{Cab06}.

\subsection{FSI in the damping model}

\begin{figure}[tp]
\begin{center}
\includegraphics[scale=0.85,  bb= 50 150 540 790]{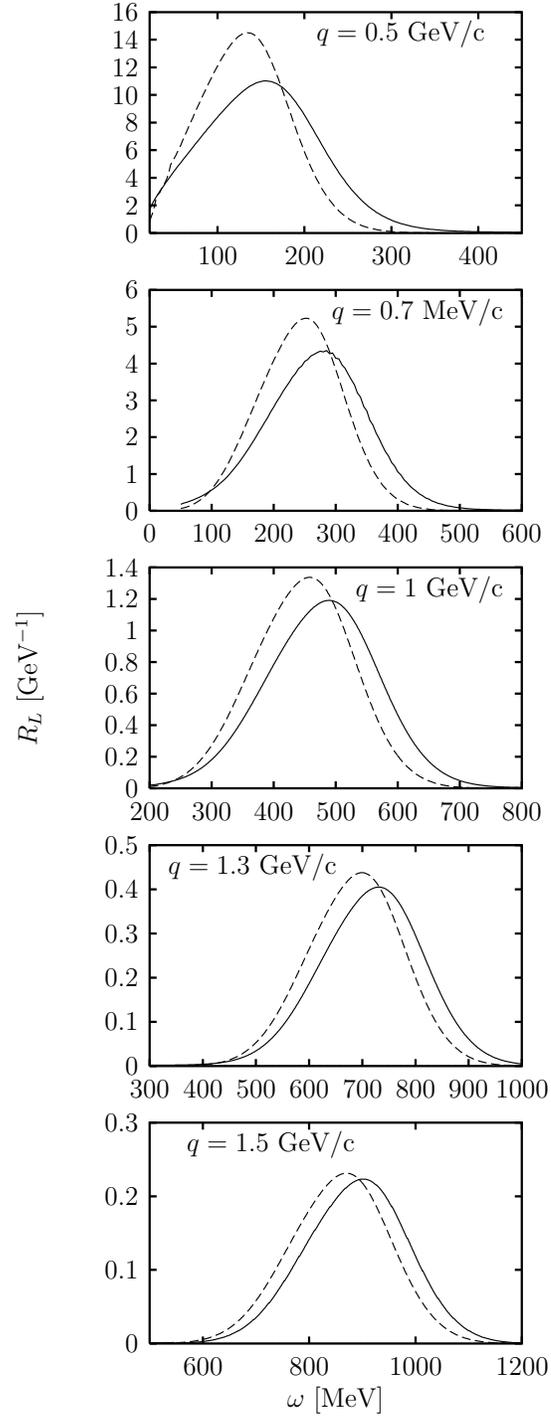}
\caption{ Effect of FSI on the longitudinal response function.
Dashed lines: results in the SR model with a Woods-Saxon potential.
Solid: including FSI in the Smith-Wambach damping model. 
 }
\end{center}
\end{figure}

\begin{figure}[tp]
\begin{center}
\includegraphics[scale=0.85,  bb= 100 570 540 780]{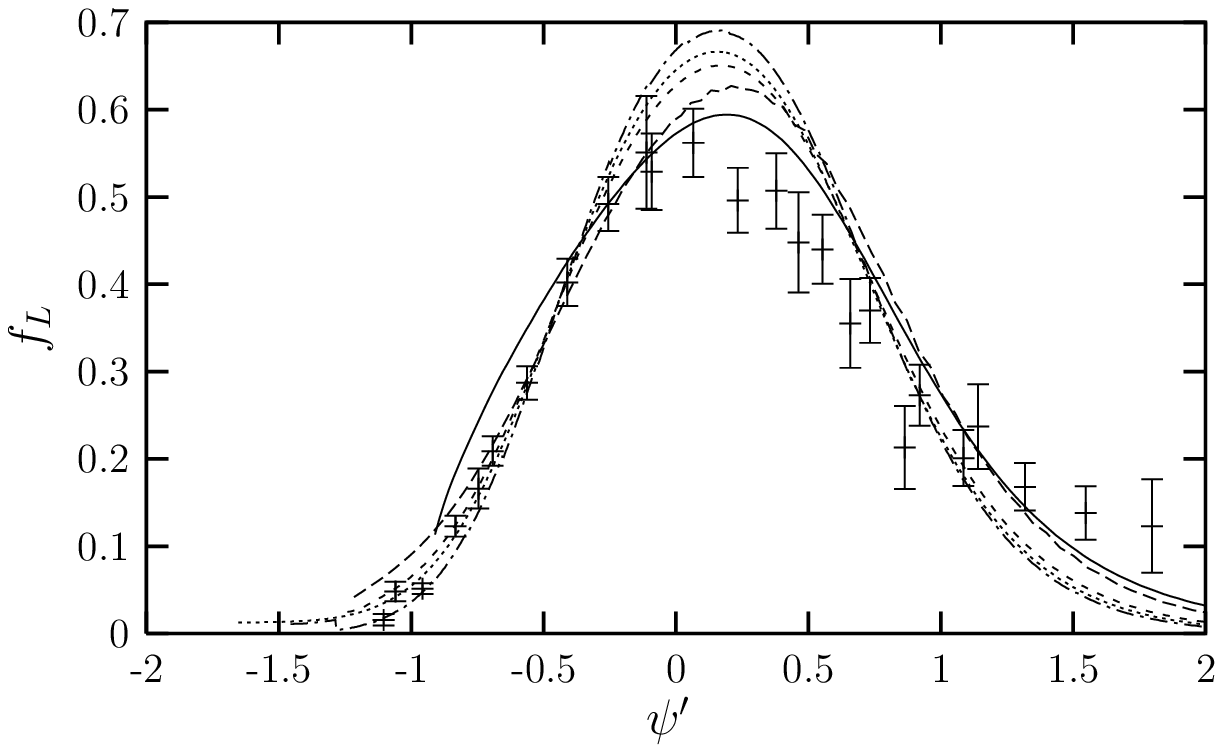}
\caption{ Longitudinal scaling function in the SR model with
Smith-Wambach damping.  Several values of the momentum transfer
$q=$ 0.5 (solid lines), 0.7 (dashed), 1.0 (short-dashed), 1.3
(dotted) and 1.5 GeV/c (dot-dashed) are plotted.  No energy shift
has been applied to the theoretical calculation.  Experimental
data from \cite{Mai02}. }
\end{center}
\end{figure}

There are two main effects embodied in the Smith-Wambach damping
model used here: a shift and a redistribution of strength, as
illustrated in Fig.~4, where we show the longitudinal response
function for five values of the momentum transfer. These effects are
a consequence of the folding integral in eq.~(\ref{folding}), which
basically produces the redistribution, and of the appearance of the
effective mass $M^*(q)$, which is responsible for the shift.
However, when the energy $\omega$ is large, the parametrization of
eqs.~(\ref{Gamma},\ref{gamma}) results in a small nucleon width:
hence the Lorentzian function $\rho(E,\omega)$ becomes close to a
Dirac delta function and only the effective mass effect remains.
This is why in Fig.~4, for the largest $q$-value, $q=1.5$ GeV/c, the
FSI just produce a shift of $R_L$. Thus, although the damping model
can produce the needed effect for low-to-intermediate momentum
transfers (i.e. redistribution of the strength as required by the
data), for higher momentum transfer this method is basically
equivalent to introducing an almost constant effective mass (of
around $M^*/m_N \simeq 0.8-0.9$).

The scaling properties of $R_L$ under the SWD model of FSI may be
seen by examining the results in Fig.~5. There one observes that the
scaling function increases with $q$, and hence that scaling of the
first kind is not respected by the model. Moreover it is almost
symmetric, whereas the experimental data, also shown in the figure,
display an asymmetry, indicating stronger re-distribution of
strength for positive  values of $\psi'$. The apparent incapability
of the SWD to describe this feature of the data should not be
considered a failure of the model, but simply an inadequate
extrapolation to high energy of a model that was originally proposed
to describe the damping of the continuum nuclear response in the
region of small energy transfer.

\subsection{The DEB approach to FSI}

\begin{figure}
\begin{center}
\includegraphics[scale=0.7,  bb= 50 110 540 790]{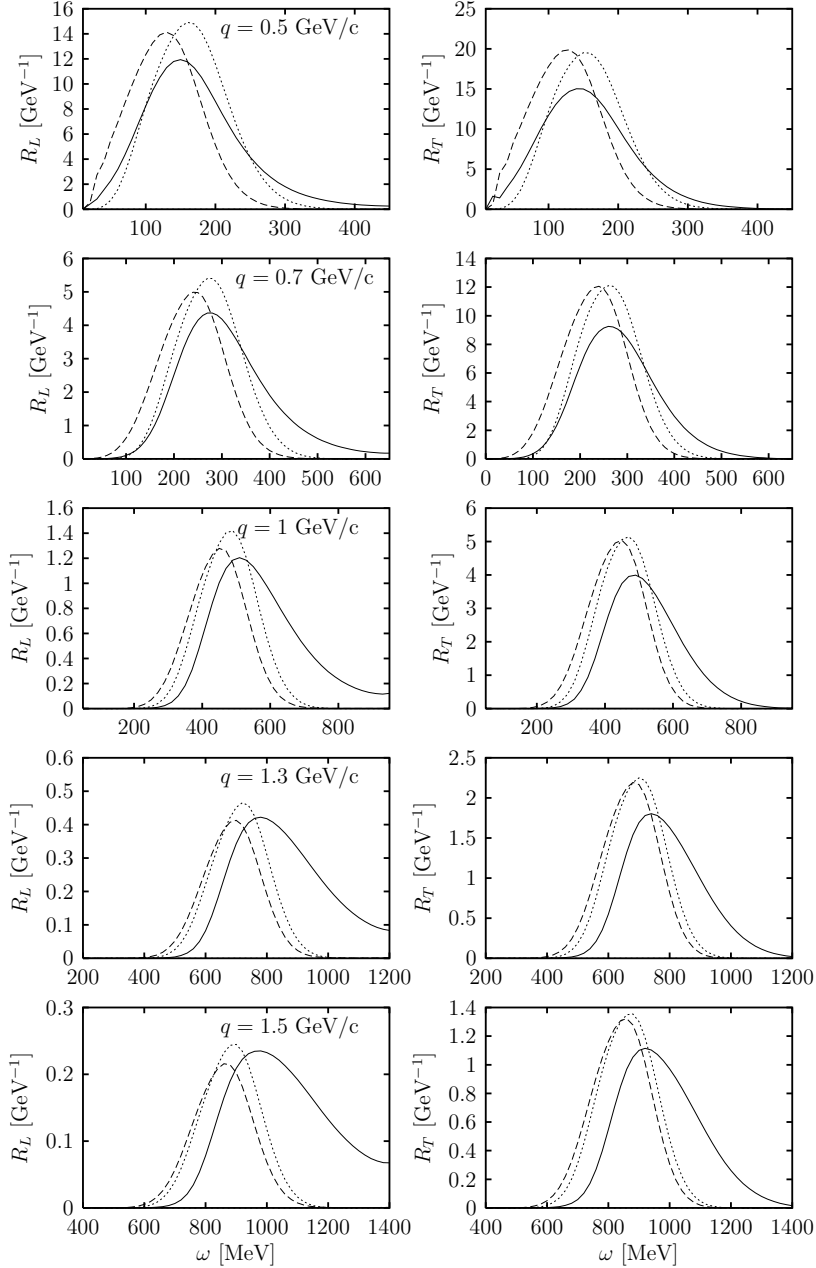}
\caption{ Effect in the DEB+D model of FSI on the L and T responses.
Dashed lines: SR-WS model. Solid: SR-DEB+D. Dotted: SR-PWIA. 
}
\end{center}
\end{figure}

The second approach to FSI in this work amounts to using the DEB
potential plus Darwin term as described above.  The interest in this
approach for intermediate-to-high energies originates from the
results found in \cite{Cab05,Cab06} where the relativistic
mean-field (RMF) model has been found to describe reasonably well
the scaling function data. In particular, the asymmetry presented by
the data appears to be obtained only in IA-based models when one
uses the same form of the (real) S and V relativistic potential in
both initial and final states. One must at this point remember that
the present focus is on inclusive reactions: one should not be
tempted to apply the optical potential fit to elastic data to
inclusive reactions such as $(e,e')$ or $(\nu_l,l^-)$, where
different channels are open \protect\cite{Horikawa80}. It is not the
purpose of this paper to find rigorous theoretical arguments to
justify the use of the Hartree potential as the best choice in
relativistic calculations. Developments along these lines will be
discussed elsewhere \cite{Udi06}. We just mention here that the
asymmetry found in the scaling function computed in the RMF is a
consequence of the large scalar and vector parts of the potential
used in the final state, which are both much smaller in the case of
the real parts of typical optical potentials or in other similar
approaches.

The goal of this section is to show that when conveniently
implemented in the SR approach, the same relativistic Hartree
potential yields results consistent with the ones obtained within
the fully relativistic mean-field RMF model.  The electromagnetic
response functions with the DEB+D choice for FSI are shown in Fig.~6
for fixed momentum transfer in the range $q=0.5$ to 1.5 GeV/c. In
the figure we show the SR results using the Woods-Saxon potential in
the final state (dashed lines) and using the DEB+D model of FSI
(solid lines).  For comparison in the same figure we also show the
SR-PWIA 
results. As one can
see, the effect of the DEB+D model is precisely a shift and
re-distribution of the strength, typically what one expects to occur
due to FSI mechanisms. For $q=$ 0.5 GeV/c the effect is similar to
the SWD results of Fig.~4. However, for higher values of $q$ the
effect of the FSI is maintained in the T response and even increases
in the case of the L response, contrary to what happens in the SWD
model. In this way, for $q=1.5$ GeV/c the total strength is not only
re-distributed, but amplified in the longitudinal response function.
The behavior of $R_L$ for high $q$ in the high-$\omega$ tail is just
a consequence of the SR expansion used for the current operator. In
fact the time component of the vector current in eq.~(\ref{JV0}) is
proportional to the $\xi_0$ coefficient which in turn is
proportional to the relativistic factor $\kappa/\sqrt{\tau}$ in
eq.~(\ref{xi0}). The fact that this factor becomes infinite on the
light cone $\omega=q$ is just a consequence of the off-shell
properties of the SR expansion, since it was performed for on-shell
nucleons and therefore its applicability should be most appropriate
for the QE-peak region. This is not a fault of the current, since
similar anomalous off-shell effects can also be seen in
prescriptions of relativistic current \cite{Cab06}. Therefore, one
should be careful when extrapolating these operators to highly
off-shell conditions, and the results of Fig.~6 should be taken with
caution in the region close to $\omega=q$, in particular when $q$ is
high and non-QE effects dominate the total cross section, as there
probably $R_L$ cannot be extracted experimentally.  In fact the
experimental data for the scaling function shown in this paper were
taken from world data in the range $q=0.5$ to 1 GeV/c.

\begin{figure}
\begin{center}
\includegraphics[scale=0.7,  bb= 50 500 540 790]{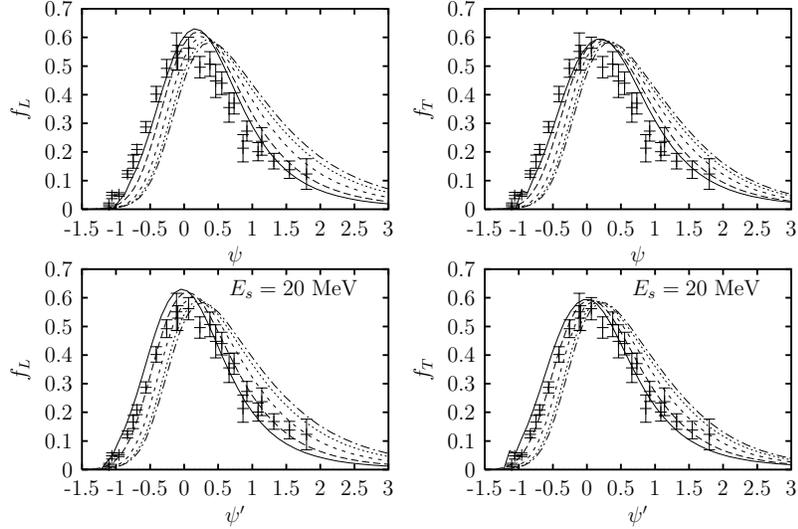}
\caption{ First-kind scaling in the DEB+D model. In each panel
results are shown for $q=$ 0.5, 0.7, 1.0, 1.3 and 1.5 GeV/c. First
row of panels: scaling functions versus $\psi$, namely, when
$E_s=0$. Second row: the same versus $\psi'$ with constant energy
shift $E_s=20$ MeV. Experimental data from \cite{Mai02}. }
\end{center}
\end{figure}

The interest in showing results for values of $q$ as high as 1.5
GeV/c is related to the study of the properties of scaling of the
first kind performed next. In Fig.~7 we show the scaling functions
in the DEB+D model.  They have been extracted from the results of
Fig.~6, and are displayed as functions of the unshifted ($\psi$) and
shifted ($\psi'$) scaling variables.  When nonzero, the energy shift
is 20 MeV in all cases.  The experimental data for $f_L$ are shown
for comparison. One can see that the DEB+D results follow the trend
of the experimental data, clearly showing the same asymmetric shape.
While the strength and width of the results of Fig.~7 are basically
the same as the data, the theoretical results are shifted to the
right with respect to the data.  Though small, the shift increases
with $q$.  Due to the shift, one would conclude that scaling of the
first kind is not perfect in this model. However, note that the data
do not contain values of $q$ as large as 1.5 GeV/c and that the
error bars are a consequence of a small dispersion, since scaling of
the first kind is not experimentally as good as scaling of the
second kind (i.e., independence of the nucleon number $A$).
Therefore a small scaling violation as that seen in Fig.~7 may not
actually be inconsistent with the data. Another interesting
conclusion from these results is that to a high degree the DEB+D
model respects scaling of zeroth kind, i.e., $f_L=f_T$, for all
$q$-values.

\begin{figure}[bt]
\begin{center}
\includegraphics[scale=0.9,  bb= 60 650 520 790]{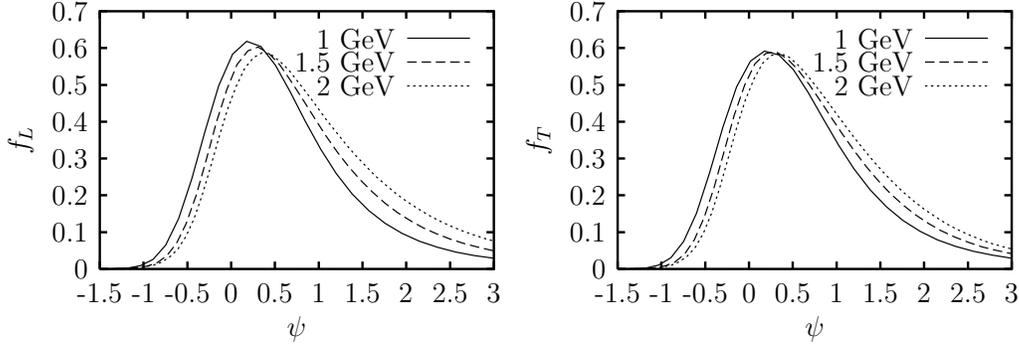}
\caption{ First-kind scaling in the DEB+D model. In each panel the
results are shown for incident electron energies $\epsilon_e=1, 1.5$
and 2 GeV. The electron scattering angle is $\theta_e=45^{\rm o}$, 
and  $E_s=0$.  
}
\end{center}
\end{figure}

Another study of scaling of the 1st kind within the DEB+D model
has been performed with results at a different kinematical setting
shown in Fig.~8. This time we show the scaling functions computed
for several incident electron energies and fixed electron
scattering angle $\theta_e=45^{\rm o}$. Here the momentum transfer
is not constant and changes with $\psi$. As a consequence, the
scaling works better and the curves are closer to each other.

\begin{figure}[tb]
\begin{center}
\includegraphics[scale=0.7,  bb= 60 380 520 790]{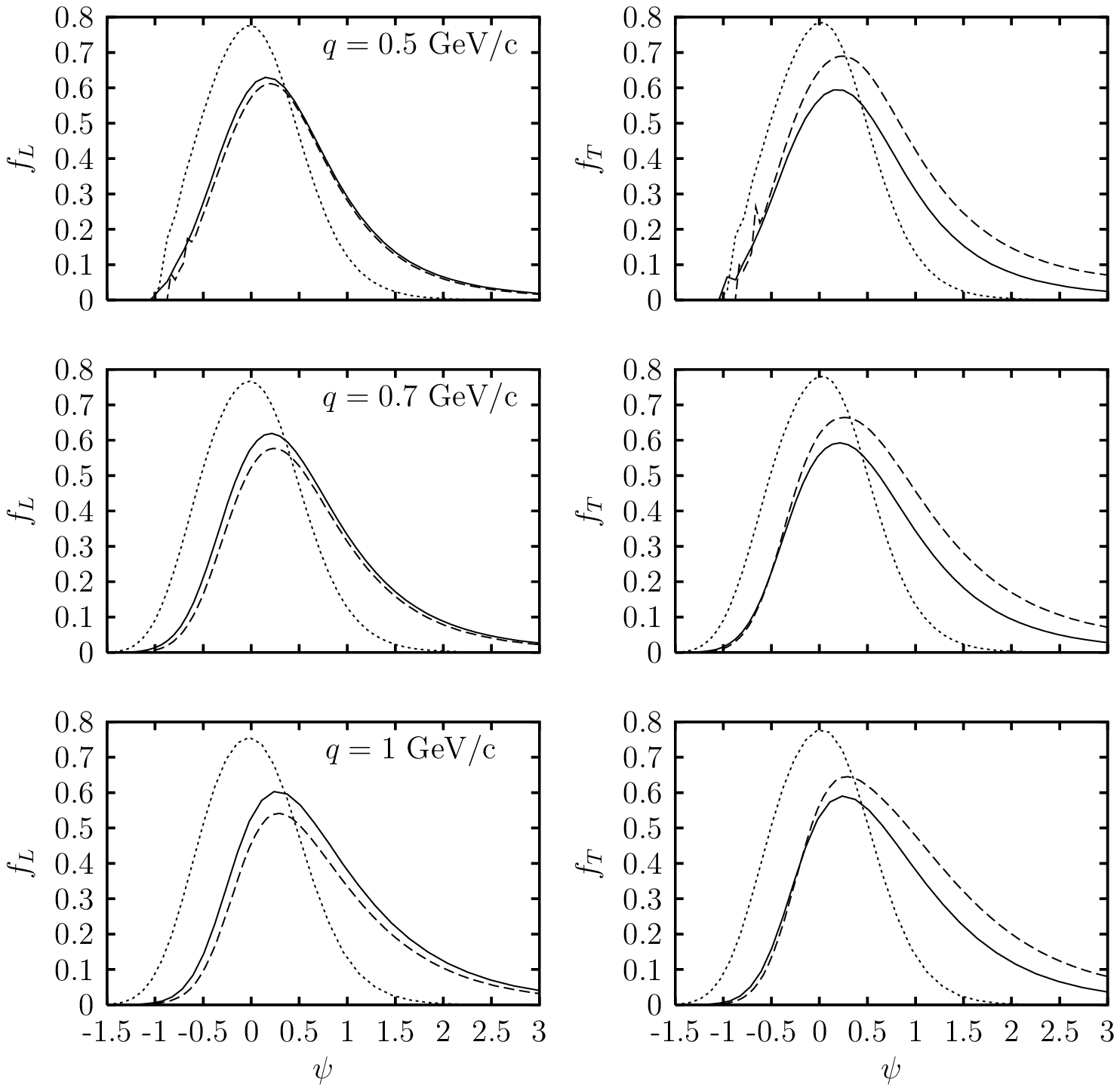}
\caption{ Electromagnetic scaling functions for three values of the
momentum transfer. Dotted lines: SR model with a Woods-Saxon
potential in the final state. Solid:  SR model using the DEB+D
approach for FSI. Dashed: RMF with the CC2 current operator. 
Here $E_s=0$. 
}
\end{center}
\end{figure}

\begin{figure}[tb]
\begin{center}
\includegraphics[scale=0.8,  bb= 60 640 520 790]{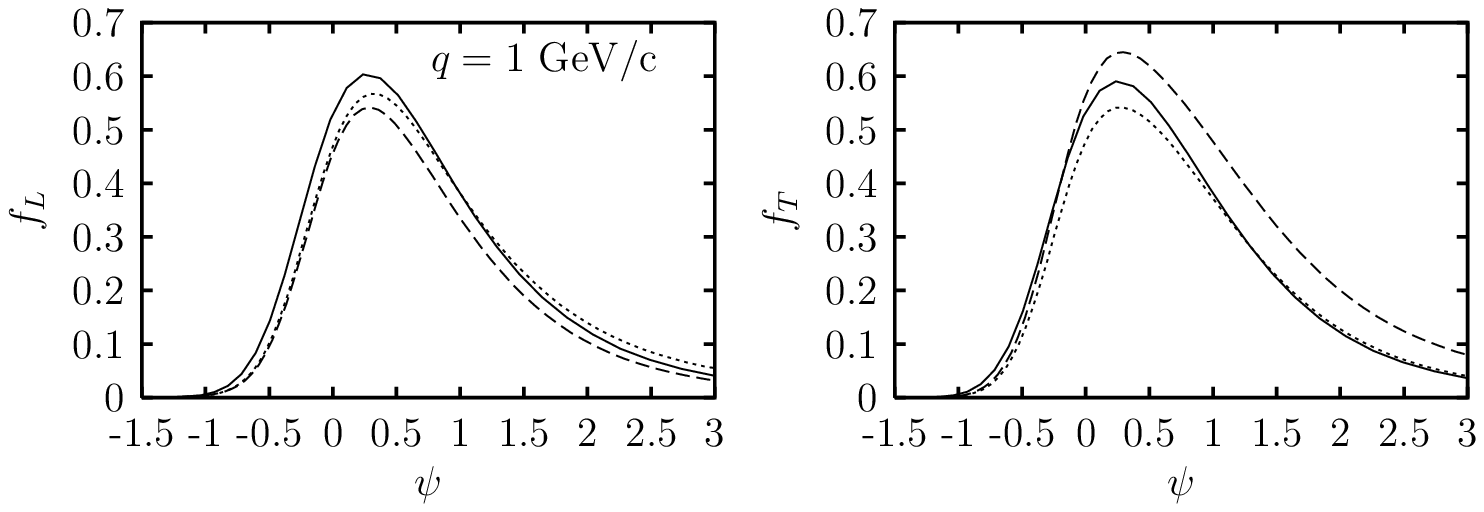}
\caption{ Electromagnetic scaling functions for $q=1$ GeV/c. Solid
lines:  SR model within the DEB+D approach for FSI. Dashed:  RMF
with the CC2 current operator. Dotted: RMF with the CC3 current
operator. }
\end{center}
\end{figure}

At this point a test of the present SR model is possible by
comparing with a fully relativistic model.  Since the final states
in the DEB+D approach are computed essentially as in the RMF
approach presented in \cite{Cab05,Cab06}, both models should give
similar results if the relevant relativistic corrections are
included properly in the SR approach. A first comparison is
performed in Fig.~9 for three values of the momentum transfer. The
RMF results are computed with the CC2 current operator.  The SR
model using the Woods-Saxon potential in the final state is also
shown for comparison. It is apparent from the figure that the
behavior of the SR-DEB+D results is similar to that in the fully
relativistic approach.  The differences seen between the two models
can partially be attributed to the peculiar off-shell properties of
the CC2 and SR currents, as well as to the four-component structure
of the wave functions involved in the RMF model compared with the
SR-DEB+D. In fact differences of the same order can also be found in
the RMF approach between results obtained using other current
prescriptions. An example is shown in Fig.~10, where we compare our
results for $q=1$ GeV/c with the RMF results employing two current
prescriptions. For this value of the momentum transfer, the DEB+D
model gives similar results to those for the RMF model with the CC3
current operator, although not exactly the same. From the results of
Figs.~9 and 10 we conclude that our model contains the relevant
relativistic corrections in this kinematical region. Our current
operator could then be considered at this level as another
prescription for off-shell extrapolation of the relativistic nucleon
current.  At this point it is also interesting to note that the RMF
model with the CC2 operator leads to a clear violation of
zeroth-kind scaling (likewise for CC1, but not for CC3), contrary to
what happens with SR-DEB+D.

\begin{figure}[tb]
\begin{center}
\includegraphics[scale=0.8,  bb= 60 360 520 790]{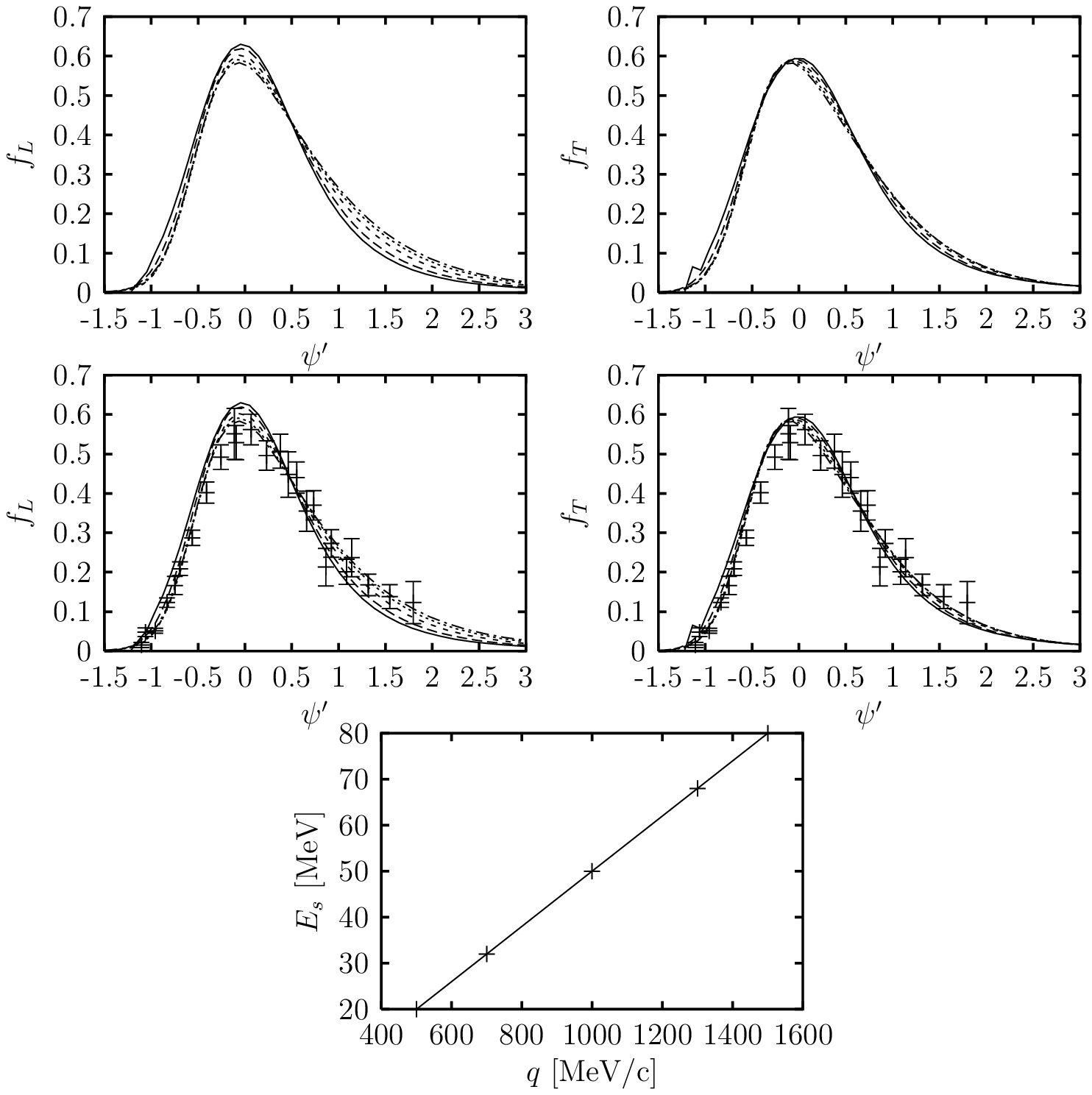}
\caption{ First-kind scaling in the SR-DEB+D model. Results are
shown for $q=$ 0.5, 0.7, 1.0, 1.3 and 1.5 GeV/c. First row: Scaling
functions versus $\psi'$ using a $q$-dependent energy shift
$E_s(q)$. Second row: the same compared with the experimental data.
Last panel: the linear behavior of $E_s(q)$. The fitted values of
$E_s$ are 20, 32, 50, 68 and 80 MeV. Experimental data from
\cite{Mai02}. }
\end{center}
\end{figure}

\begin{figure}[tb]
\begin{center}
\includegraphics[scale=0.8,  bb= 60 500 520 790]{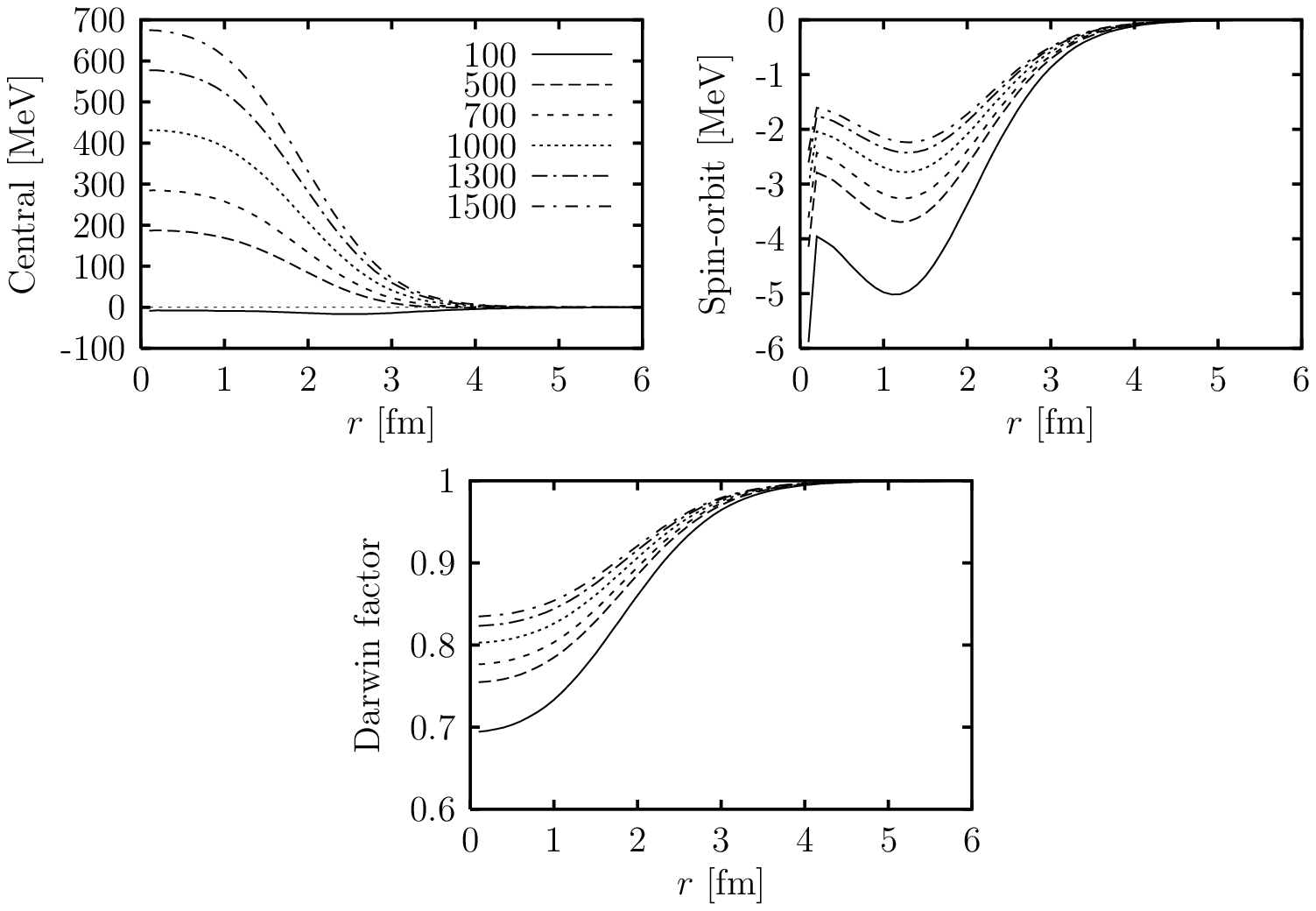}
\caption{
First row: DEB potential for final nucleon kinetic energies
 100--1500 MeV.
Last panel: the same for the Darwin factor.
}
\end{center}
\end{figure}

Motivated by the results of Fig.~7, where scaling of first kind is
fulfilled except for a small shift, and in order to explore further
the scaling properties of our model, we have performed the
calculation of the scaling variable $\psi'$ using a $q$-dependent
energy shift $E_s(q)$. The value of $E_s(q)$ is fitted to get the
maximum of $f_L$ at $\psi'=0$.  The results of this fit are shown in
Fig.~11.  One sees that the collapse of the curves is improved by
this procedure. Although the scaling is not perfect, it is
remarkable that for a wide range of $q$ values ($q=0.5$ to 1.5
GeV/c) our results give essentially the same scaling function. The
scaling is better for the transverse function $f_T$, while the width
of $f_L$ slightly increases with $q$.  In Fig.~11 we also show a
comparison with the experimental data.  The agreement between theory
and experiment is excellent.  We recall that data refer only to the
analysis of the longitudinal response, hence caution should be
exercised when comparing the calculated transverse scaling function
$f_T$ and data. Finally, we also present in Fig.~11 the shift energy
$E_s(q)$ as a function of $q$, showing that the dependence of the
fit is clearly linear. This linear behavior of $E_s(q)$ is connected
to the energy dependence of the DEB potential, which is also linear
\cite{Udi95,Udi01}.  We illustrate this dependence in Fig.~12, where
we plot the central and spin-orbit parts of the DEB potential for a
range of final nucleon kinetic energies from 100 to 1500 MeV.  More
details on the properties of the DEB potential can be found in
\cite{Udi95}. From Fig.~12 it is evident that the central part of
the DEB potential increases linearly with the energy and that it is
very repulsive for high energy.  The spin-orbit part, on the other
hand, is rather small and unimportant for these energies. In the
same figure we also show the Darwin factor $K(r,E)$ for the same
nucleon energies.

\begin{figure}[tbp]
\begin{center}
\includegraphics[scale=0.65,  bb= 60 100 520 790]{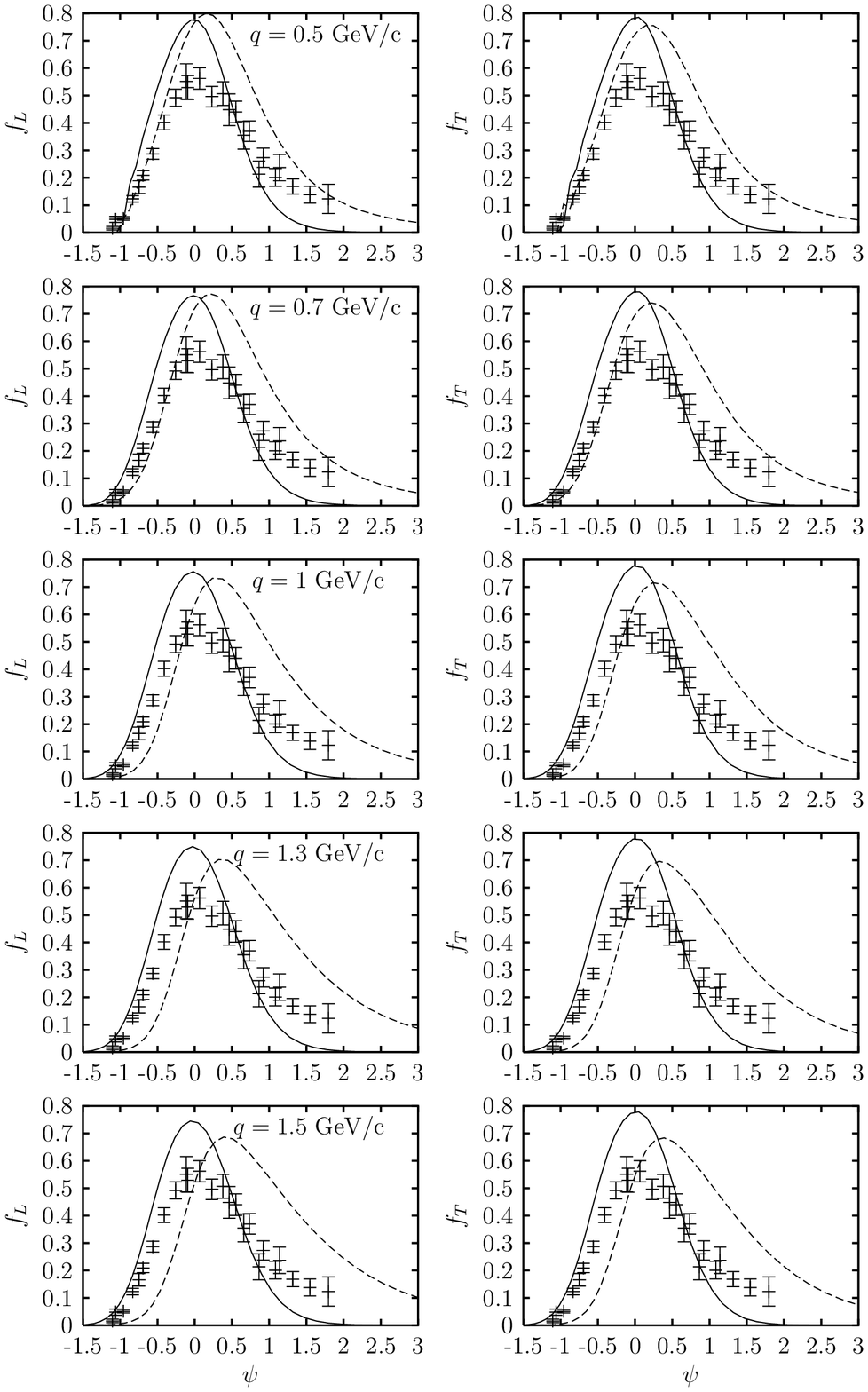}
\caption{ Effect of the DEB potential without the Darwin term. Solid
lines: SR model with the same Woods-Saxon potential in the initial
and final states.  Dashed: SR-DEB results with the Darwin term set
to one. Experimental data from \cite{Mai02}. 
}
\end{center}
\end{figure}

The highly repulsive behavior of the DEB potential is the main
reason for the strength re-distribution seen in our results. A
repulsive potential favors the emission of nucleons having high
enough available energy. Accordingly, more strength is placed in the
high energy tail of the cross section or, equivalently, at positive
values of $\psi'$ in the scaling function.  One should mention here
that the DEB potential alone is not enough to produce reasonable
results for the scaling function. The Darwin factor, $K(r,E)$, shown
in the last panel of Fig.~12, must also be applied to the wave
function. We illustrate this point in Fig.~13, where we compare the
SR-WS model (namely with Woods-Saxon potential in the initial and
final states) with the SR-DEB, but with the Darwin term set to one.
One can see that the DEB potential alone produces the mentioned
shift and widening of the scaling function to higher energies, due
to the repulsive character of the potential, although the strength
is still too high when compared with the data.

\begin{figure}[tbp]
\begin{center}
\includegraphics[scale=0.65,  bb= 60 100 520 790]{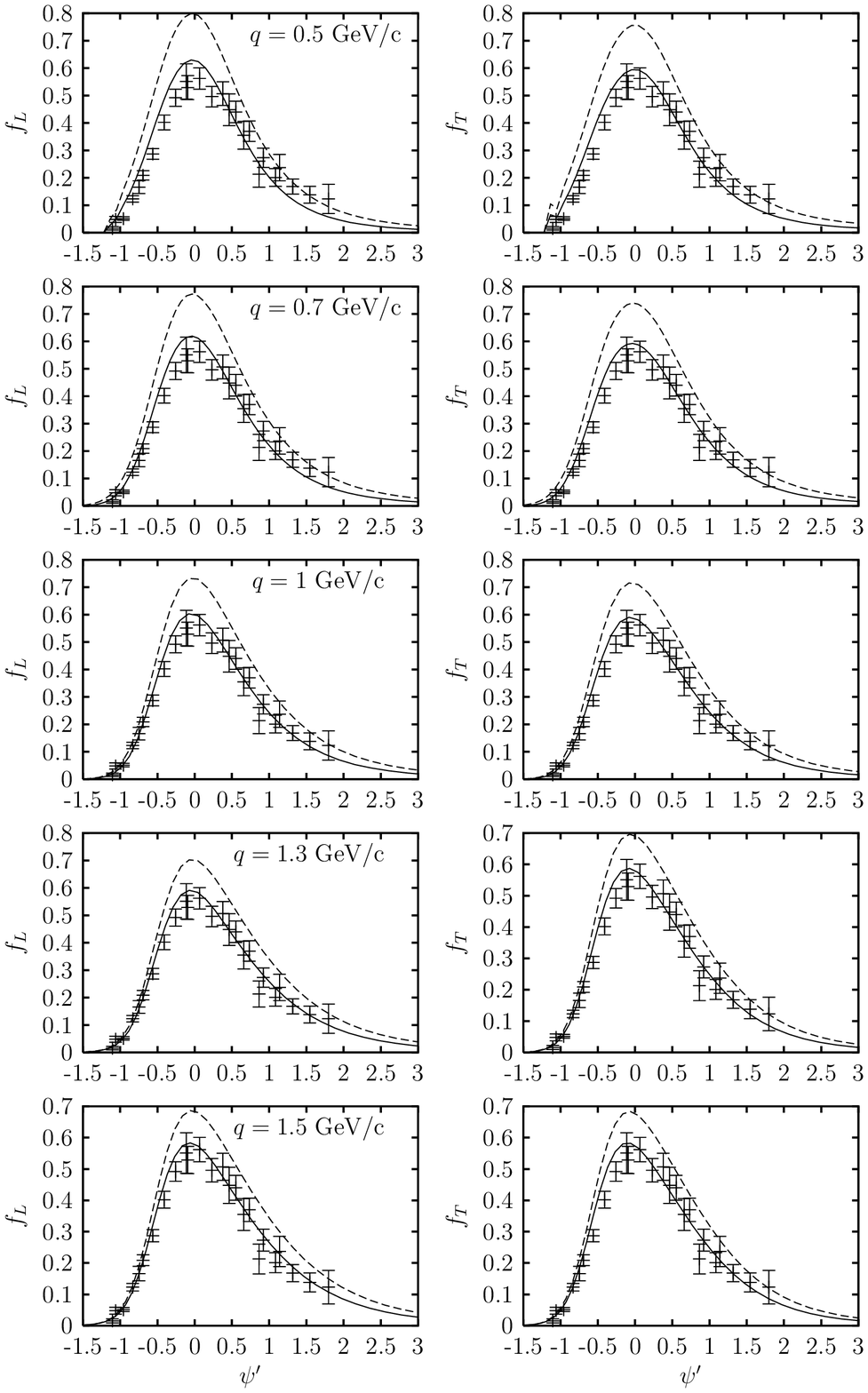}
\caption{ Effect of the Darwin term on the scaling function in the
SR model. Solid: total DEB+D result. Dashed: results with the Darwin
term set to one. Experimental data from \cite{Mai02}. 
}
\end{center}
\end{figure}

The effect of the Darwin term can be appreciated in Fig.~14, where
we plot the scaling functions in the SR-DEB approach with and
without that factor (or equivalently by setting $K(r,E)=1$ in the
calculation). Since $K(r)<1$ inside the nucleus, its effect is a
reduction of the scaling function. This reduction is precisely the
one needed to reach the experimental data, also shown in Fig.~14.
Thus the Darwin factor is essential in this approach to FSI.

One should mention that the full DEB+D model is not equivalent to
solving a Schr\"odinger equation with a Hermitian local potential.
Actually, the wave function in eq.~(\ref{psiup}) is a solution of
such an equation with a non-local potential, since apart from the
dependence on the energy $E$, dependence on the momentum or
gradient operator also appears from the reduction of the Dirac
equation \cite{others,jaminon,Udi95,Udi01}. This means that the
solution of the Schr\"odinger equation with DEB potential plus
Darwin factor is not comparable to any other non-relativistic or
relativized solutions obtained with {\em local} potentials.  The
non-locality contained in the DEB+D approach introduced here, also
implicit in the Dirac equation and made explicit by means of the
DEB+Darwin reduction, thus becomes essential in producing the kind
of effects that usually appear in non-relativistic approaches of
FSI related to the role of correlations and/or exchange terms in
the optical potentials.  In particular, one could argue that in
the SWD model discussed above, the inclusion of 2p-2h intermediate
states gives rise also to non-local effects in the final state.

\begin{figure}[tbp]
\begin{center}
\includegraphics[scale=0.65,  bb= 60 100 520 790]{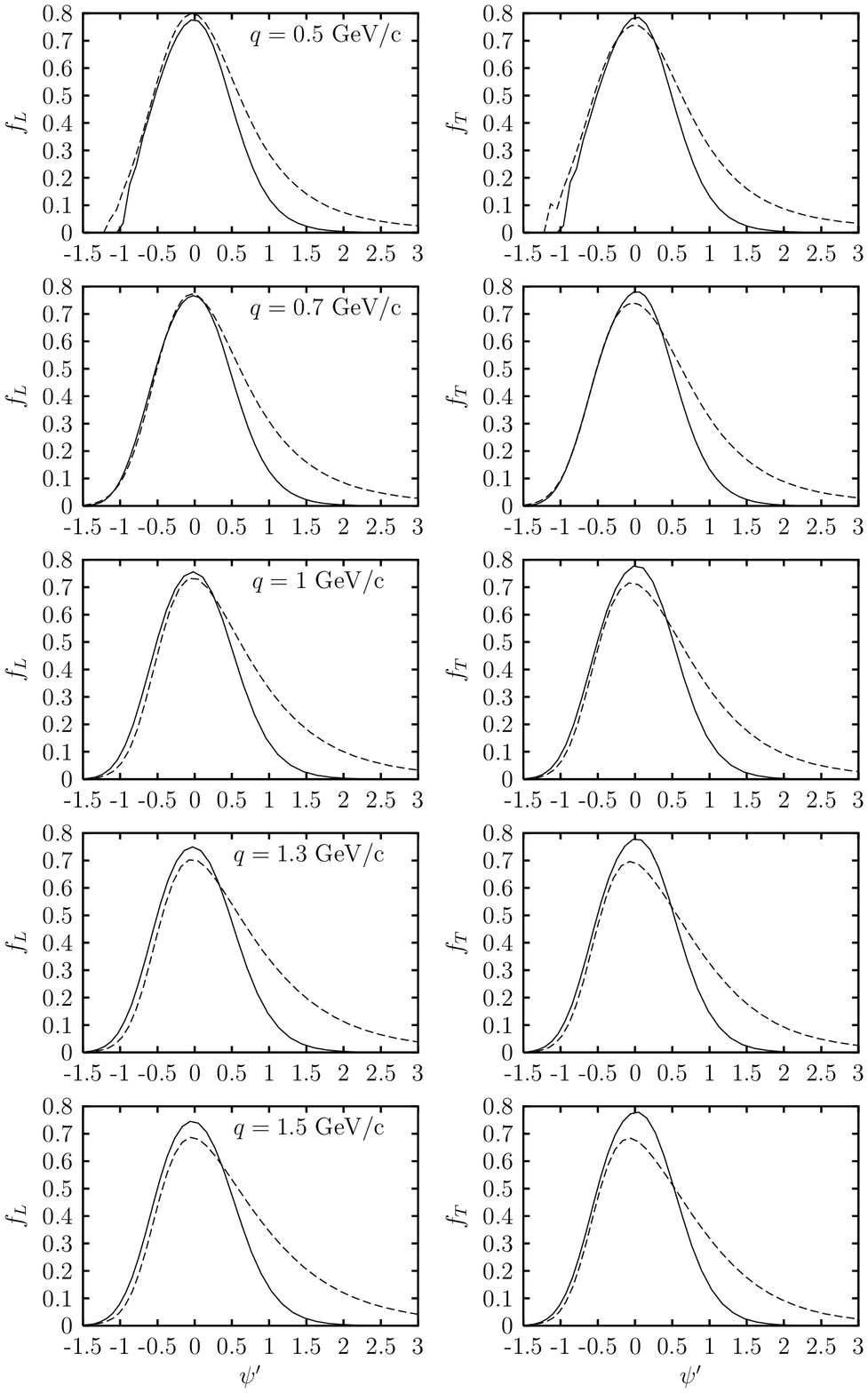}
\caption{ Effect of the DEB potential without the Darwin term. Solid
lines: SR model with the same Woods-Saxon potential in the initial
and final states.  Dashed: SR-DEB results with the Darwin term set
to one and with the $q$-dependent energy shift $E_s(q)$. 
}
\end{center}
\end{figure}

A better appreciation of the effect of the DEB potential without the
Darwin factor can be seen in Fig.~15, where we compare the scaling
functions in the SR-WS and SR-DEB potentials. Once again the Darwin
factor is set to one. The $q$-dependent shift $E_s(q)$ has been
applied to the DEB results. The widening of the distribution to
higher energies, producing a longer tail, is clear.  Note also that
the DEB potential does not modify the results in the region
$\psi'<0$, and that there is a reduction of the maximum in the
region $\psi'\sim0$ (the magnitude of the reduction increases with
$q$).

\subsection{Applications to neutrino reactions}

\begin{figure}[tbp]
\begin{center}
\includegraphics[scale=0.65,  bb= 60 110 520 800]{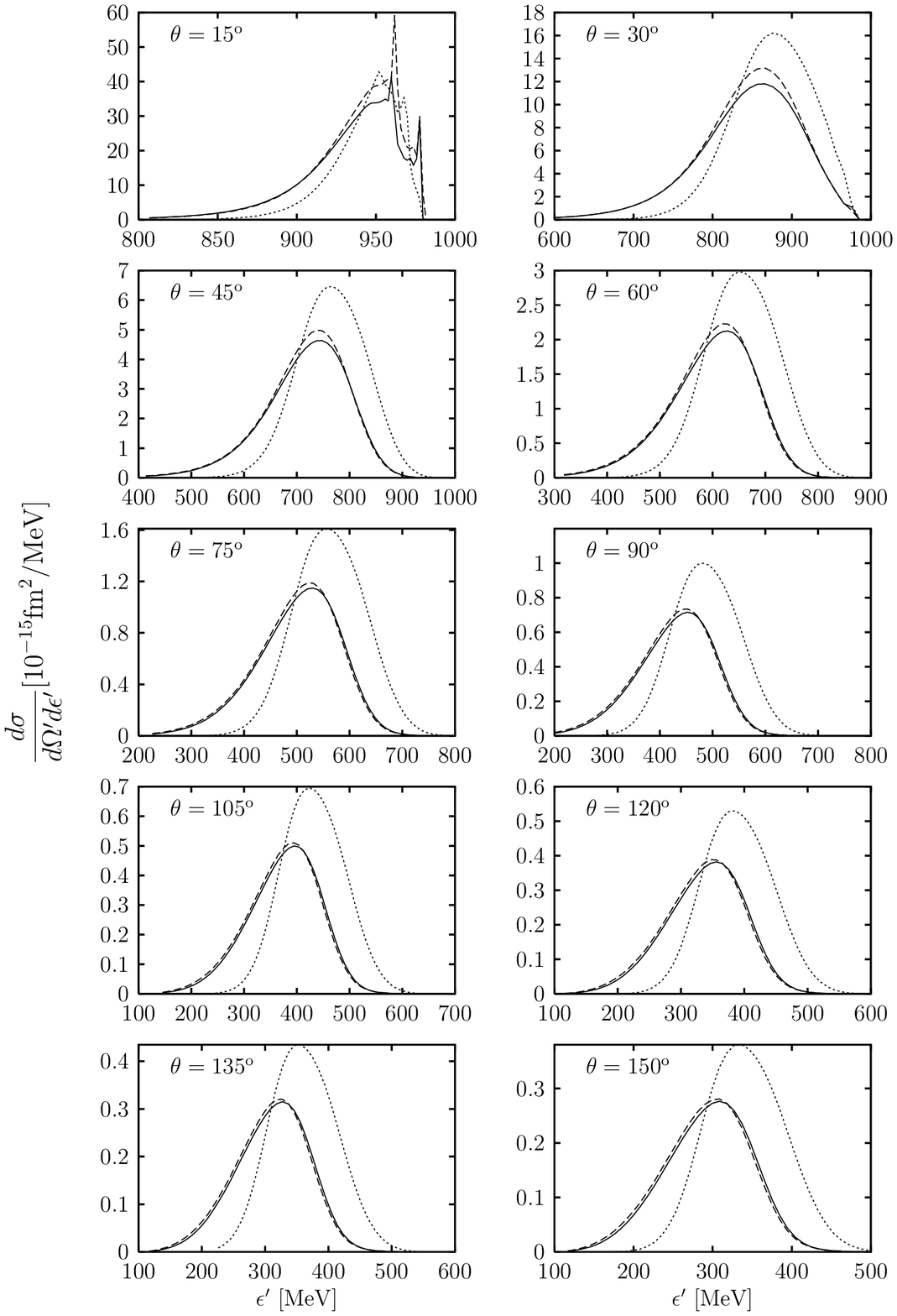}
\caption{ Double-differential cross section for the
$^{12}$C$(\nu_\mu,\mu^-)$ reaction for incident neutrinos of 1
GeV, computed with several versions of the SR model. Dotted:
Wood-Saxon potential. Solid: DEB+D.  Dashed: reconstructed from
the $(e,e')$ DEB+D model using the SuSA. }
\end{center}
\end{figure}

\begin{figure}[tbp]
\begin{center}
\includegraphics[scale=0.65,  bb= 60 110 520 800]{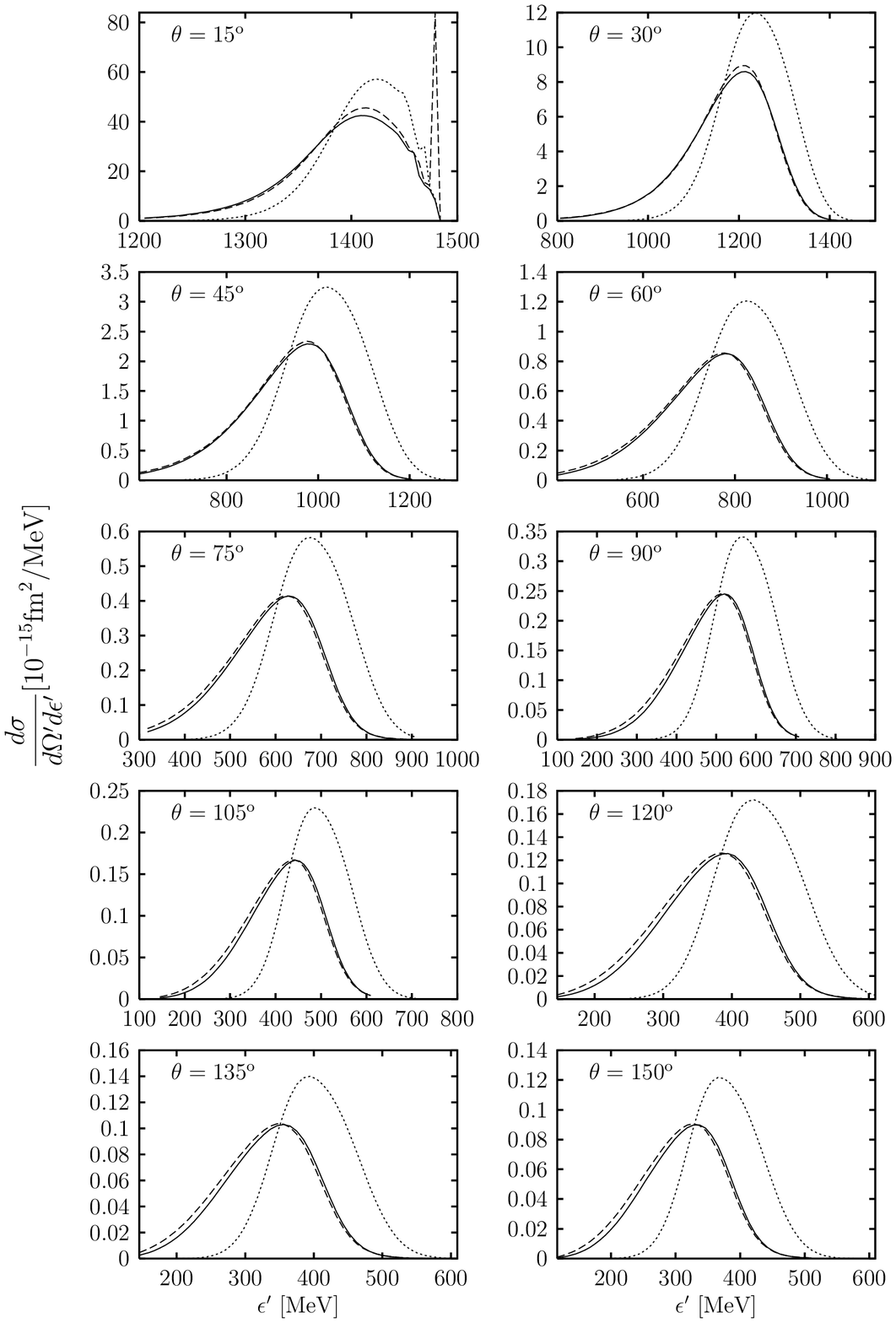}
\caption{ The same as Fig.~16, but for neutrinos of 1.5 GeV. }
\end{center}
\end{figure}

We now exploit the SR-DEB+D model introduced in the last section to
make predictions for QE CC neutrino reactions.  Given the success of
that model in describing the $(e,e')$ superscaling data, the
neutrino cross section results obtained with our model can be
considered at least reasonable in a kinematical range from
intermediate-to-high energies and momentum transfers.

A selection of results for the $^{12}$C$(\nu_\mu,\mu^-)$
differential QE cross section are presented in Figs.~16 and 17,
for incident neutrino energies of 1 and 1.5 GeV, respectively. We
show results for a range of scattering angles from $15^{\rm o}$ to
$150^{\rm o}$. All of the results in the figures have been
obtained with the SR model with different ingredients for the FSI.
The results within the DEB+D model of FSI are shown with solid
lines, while the dotted lines are the SR-WS results using the same
Woods-Saxon potential in initial and final states.  We display the
cross section as a function of the final muon energy $\epsilon'$.
Therefore the tail produced by the FSI can now be seen to the left
of the peak of the SR-WS results. Finally, with dashed lines we
show the results obtained in a third calculation which closely
follows the general procedure originally introduced in
\cite{Ama05b} and referred to as SuSA approach.

Just as when the SuSA approach was performed using data, here the
theoretical $(e,e')$ cross section is divided by the single-nucleon
factor to obtain the scaling function, which in turn is multiplied
by the relevant CC weak factor, to obtain the neutrino cross
section. That is, this procedure is similar to the SuSA approach in
\cite{Ama05b} except that instead of using the ``experimental''
superscaling function obtained from the analysis of $(e,e')$ world
data, in this case we make use of the scaling function derived from
$(e,e')$ results with the SR-DEB+D model. Notice that the
reconstruction of the $(\nu_\mu,\mu^-)$ cross section from the
$(e,e')$ data (SuSA) or $(e,e')$ calculations (we denote it as
superscaling-based approach to avoid confusion with SuSA which
implies the ``experimental'' scaling function) assumes that the
scaling function for both reactions is similar for the kinematics of
interest.
Thus in Figs.~16 and 17, we treat the solid lines of the DEB+D model
as ``exact'' and check the accuracy of the superscaling-based
approach by comparison of the dashed lines with the solid ones. One
can see that in general the superscaling-based results are very
close to the DEB+D results. The differences between the two sets of
results become smaller as the scattering angle and neutrino energy
increase, since bigger values of the momentum transfer are involved.
The biggest differences appear for very small angles, implying
rather small values of the momentum transfer. In these cases the
description of the giant resonance region needs a nuclear model
containing collective degrees of freedom such as the RPA, not
included in our approach. An example is shown for the case of
$\theta=15^{\rm o}$ in Figs.~16 and 17. The presence of the narrow
peaks appearing as potential resonances clearly violates the scaling
of the cross section in that region. However, for higher values of
the scattering angle the superscaling-based analysis begins to be
applicable, and its predictions can be considered quite reasonable,
with an error typically below 10\%.

\section{Conclusions}

In this work we have presented improvements over the
semi-relativistic approach to electron and neutrino reactions for
intermediate-to-high energies and momentum transfers in the QE
region. Going a step beyond the continuum shell model, we have
explored two approaches to describe the FSI: the Smith-Wambach
damping model and the Dirac-equation-based potential. The former is
applicable to low energies and has simply been extrapolated here to
relativistic energies, while the latter is more appropriate for the
kinematical regime considered in this paper and has been our main
focus.

Thus, in addition to using the semi-relativistic expansion of the
nucleon current in powers of $p/m_N$ and relativistic kinematics,
we have used the DEB form of the relativistic Hartree potential
term to describe the continuum wave function of the ejected
nucleon. Furthermore, we have included in the wave function the
non-localities arising from the reduction of the Dirac equation by
multiplying it by the corresponding Darwin factor.

Firstly, we have focused on the analysis of the electromagnetic
response functions. In particular we have investigated the
properties of scaling of the first kind displayed by our results.
The longitudinal and transverse scaling functions have been computed
for a wide range of momentum transfers, from 0.5 to 1.5 GeV.
Several aspects and details of the different theoretical ingredients
embodied in our model have been analyzed. In particular, we have
examined the effects of the relativistic corrections and we have
presented comparisons with fully relativistic results.

We have found that our model approximately fulfills scaling of the
first kind except for a small energy shift, $E_s(q)$, which turns
out to be linear in $q$. This behavior has been connected to the
repulsive character of the DEB potential, which also depends
linearly on the nucleon energy. The Darwin factor has been found
to be essential for the description of the experimental scaling
function data. The Darwin factor is needed to correct for
mathematically unavoidable non-localities arising in the
differential equation describing the upper component of the
relativistic nucleon wave function.

Is is a remarkable result of this work that, except for the energy
shift, our model gives essentially the same scaling function for a 1
GeV wide range of three-momenta $q$ --- spanning a kinematical
region that extends from non-relativistic to relativistic conditions
--- and that the behavior of the theoretical scaling function is
essentially the same as that of the experimental data. The study of
the theoretical energy shift performed here can be of help for
future analyses of the scaling properties of experimental data. In
particular, it suggests that the small scaling violations of the
first kind found in the $(e,e')$ data can be used to extract
valuable information about the strength of the FSI.

Finally, we have presented an application of the model to the
inclusive CC neutrino reaction $^{12}$C$(\nu_\mu,\mu^-)$ in the
region of the QE peak. Results have been presented in a range of
kinematics for several neutrino energies and muon scattering angles.
In particular, we have used  our model to investigate theoretically
the kinematical range of validity of the superscaling approach in
reconstructing the neutrino cross section from the electromagnetic
scaling function. The validity of the superscaling-based approach
has been verified with our model for kinematics involving values of
the momentum transfer large enough such that the QE region is not
contaminated by the presence of giant resonances. For incident
neutrinos of 1 GeV, this happens for angles typically bigger than
$\sim$$15^{\rm o}$. In these cases, within our model the
superscaling-based approach is seen to be very successful in being
able to reconstruct the neutrino cross section from the
electromagnetic one.

\section*{Acknowledgments}

This work was partially supported by funds provided by DGI (Spain) and
FEDER funds, under Contracts Nos. BFM2002-03218, FIS05-01105,
FPA2005-04460, FPA2006-13807 and FPA2006-07393, by the Junta de Andaluc\'{\i}a,
by the Comunidad de Madrid and UCM,
and by the INFN-CICYT collaboration agreement (project
``Study of relativistic dynamics in electron and neutrino scattering'').
It was also supported in part (TWD) by the U.S. Department of Energy
under cooperative research agreement No. DE-FC02-94ER40818.
We thank the support of INT Program 06-2b,
``Neutrino Response Functions from Threshold to 10 GeV'',


\end{document}